# Continuous wave superconducting radio frequency electron linac for nuclear physics research

Charles E. Reece

Thomas Jefferson National Accelerator Facility, Newport News, VA 23606 USA

(Received: 15 September 2016)

*Abstract*

CEBAF, the Continuous Electron Beam Accelerator Facility, has been actively serving the nuclear physics research community as a unique forefront international resource since 1995. This CW electron linear accelerator (linac) at the U.S. Department of Energy's Thomas Jefferson National Accelerator Facility (Jefferson Lab) has continued to evolve as a precision tool for discerning the structure and dynamics within nuclei. Superconducting RF (SRF) technology has been the essential foundation for CEBAF, first as a 4 GeV machine, then 6 GeV, and currently capable of 12 GeV. We review the development, implementation, and performance of SRF systems for CEBAF from its early beginnings to the commissioning of the 12 GeV era.

## I. INTRODUCTION

The core mission of Jefferson Lab is research to understand how the nucleon's behavior when interacting with other particles changes from that of an independent entity to that of three interacting quarks. Here we address the story of how SRF technology was brought into service of this mission in CEBAF and how it has served the evolution of CEBAF over the past 30 years.

There are two key application drivers that push one to consider employing the complexity of continuous wave (CW) superconducting radio frequency (SRF) technology. The first is beam energy precision. It is much easier to create very stable beam energy and very low energy spread with CW resonant systems rather than pulsed systems. Multiple feedback loops in the low level RF controls allow the creation of beams with high precision energy definition. The second is operating cost. As the required integrated acceleration voltage increases, the pressure to keep real estate footprint low grows rapidly and thus pressure to increase active acceleration gradient does as well. CW acceleration gradients of 5 MV/m, and now much higher, are simply not sustainable apart from the use of superconductors, otherwise, resistive wall losses become nearly impossible to control thermally, and the associated electric power bill becomes prohibitively high.

From the mid 1970's nuclear physics interests had clearly established a need for beams of multi-GeV electrons with which to probe nuclear structure with precision. There was the need to support coincidence experiments with meaningful event rates, very narrowly discriminated kinematics with less than one event per electron bunch on target. This need encourages a solution with very high bunch repetition rate and relatively low charge per bunch. While the specific frequency choice was rather arbitrary and dictated by particular history, in the case of CEBAF, a bunch repetition rate of 499 MHz was chosen to work just fine. A fundamental RF frequency of 1497 MHz then allows for three simultaneous bunch trains serving three independent experimental halls, each bunch train having independent current amplitude.

## II.   KEY PARAMETERS FOR CEBAF

In CEBAF's case, further cost optimization led to adoption of a recirculating linac concept, wherein the same CW RF linac voltage is used multiple times to accelerate the relativistic electron beam. [1] This beam recirculation concept had been well demonstrated in the RLA at HEPL, Stanford University.[2] CEBAF was first designed as a 4-pass recirculating linac. After the tunnel architecture was set and under construction, the decision was taken to change to a 5-pass design because adding the additional separation, arc, and recombiner magnet systems cost less than 20% of the baseline linac costs. The most basic beam characteristics, listed in Table 1, were established early.

Table 1: CEBAF initial beam requirements

| | |
|---|---|
| Electron energy, E [GeV] | $0.5 \leq E \leq 4.0$ (upgradable to 6.0) |
| Average current [μA] | 200 |
| Transverse emittance (95%, 1 GeV) [m] | $2 \times 10^{-9}$ |
| Energy spread (95%) | $1 \times 10^{-4}$ |
| Duty factor | 100% |
| Simultaneous beams | 3 |
| Simultaneous energies | $\leq 3$ |

## III.   INFLUENCES ON THE LAUNCH OF CEBAF

The institutional history associated with the founding of CEBAF is an interesting story at the level of national science policy, regional and inter-laboratory politics, and technology vision. Catherine Westfall documented this history in 1994 [3],  and Franz Gross has summarized the nuclear physics context for the creation of CEBAF.[4] A series of National Academy of Sciences and Nuclear Science Advisory Committee reports identified and reaffirmed that a powerful, precision electron accelerator was the top priority for nuclear physics research in the US.

The Vogt Subcommittee of the Nuclear Science Advisory Committee stated the physics objective of this new accelerator:

"The search for new nuclear degrees of freedom and the relationship of nucleon-meson degrees of freedom to quark-gluon degrees of freedom in nuclei is one of the most challenging and fundamental questions of physics."[5]

A competitive design selection process eventually resolved with selection of a proposal by the Southeastern University Research Association (SURA). This proposal foresaw a 2 GeV, S-band, disc-loaded, pulsed, room temperature linac that would reach 4 GeV by one "head to tail" beam recirculation, and CW beams at the experimental end stations through the use of a stretcher ring. Soon after Hermann Grunder was appointed Director of the new effort in 1985, he commissioned a technology review which led to the brazen proposal to completely redesign the accelerator based on superconducting RF. It was judged that recent international progress with the technology provided adequate basis for the change. In addition, the availability of a 1947 MHz SRF cavity solution, only just successfully demonstrated in a test with stored beam in CESR at Cornell University in 1984, [6] for direct adoption as the building block for an SRF-based CEBAF was essential for the project to maintain its timeline and its tenuous Congressional budgetary support.

On February 13, 1987, construction started on CEBAF, a 4-GeV, 200-µA, continuous beam, electron accelerator facility designed for nuclear physics research. This was the largest scale application of SRF technology yet undertaken and the first research institution fully dependent on it.[7]

The very short time scale available for design finalization constrained design choices. Concern over cavity performance limitations due to surface contamination led Peter Kneisel to propose the use of hermetically assembled and tested pairs of cavities as the basic building block of the accelerator.[8] Once the required RF and vacuum performance was demonstrated at 2 K in a test dewar, the cavities could be maintained under controlled vacuum conditions throughout subsequent cryomodule assembly, installation and operation. These pairs were enveloped fully in individual stainless steel helium vessels and the helium vessels each built out to a "cryounit". Four cryounits assembled together and bounded by supply and return endcans constituted one "cryomodule."

To accomplish this architecture required the use of a ceramic RF window on the input waveguide coupler line. This window bounded the beamline vacuum, transmitted the 5 kW klystron power driving each cavity, and was mounted directly to the cavity in the 2 K helium bath. While the hermetic pair design choice was well motivated under the cleanliness standards of the time, the choice also created an unrecognized vulnerability which later turned out to have significant impact on the maximum useful energy of CEBAF for nuclear physics research.

Numerous other design challenges were quickly engaged to realize this complex new accelerator. Christoph Leemann led the accelerator design effort, teaming with Ronald Sundelin who led the SRF science and technology effort. Leemann reported the CEBAF design overview to the SRF community at the Third Workshop on RF Superconductivity hosted by Argonne National Laboratory in 1987.[9]

## IV.  HISTORY OF CEBAF CAVITY DESIGN

In the early 1980's Maury Tigner at Cornell University proposed building a 50-on-50 GeV $e^+e^-$ collider ring, dubbed CESR-II, to exploit the physics at the $Z^0$ resonance.[10] This was envisioned to be quite cost competitive with both LEP and SLC and relied on exploitation of superconducting RF. Maury's SRF development team had been working at Cornell since 1969 supported by NSF and included Ron Sundelin, Joe Kirchgessner, Hasan Padamsee, Peter Kneisel, and Larry Phillips. [11] They developed and tested two generations of niobium 1.5 GHz structures in beam tests in the CESR ring.[12] The specific target frequency was the third harmonic of the CESR RF, 1499.2845 MHz. (The CESR RF frequency had previously been chosen so that Cornell could use PEP/PETRA 500 MHz klystrons.)

The second generation of these cavities were composed of 5 cells, elliptical in cross-section to avoid the multipacting phenomenon which had limited the performance of the HEPL cavities.[13, 14] Nothing like contemporary design codes existed at the time, with the result that the forming dies produced cavities that were somewhat low in frequency. These were mechanically tuned as necessary to support the very successful beam test at 1499.2845 MHz in 1984.[6, 15, 16] Each cavity also incorporated a reduced-height waveguide input coupler and two smaller waveguide couplers to extract beyond cut-off higher-order-mode (HOM) power. Sundelin was strongly attracted to the inherent broadband coupling provided by waveguide couplers in contrast to coaxial style RF couplers.

Even before the successful CESR beam test, it was clear that both LEP and SLC were proceeding toward construction and there was no hope that NSF would support a CESR-II at Cornell. The demonstrated cavity design appeared to be an orphan. But Hermann Grunder had a vision, and CEBAF needed the Cornell cavity – and SRF scientific staff.

Because CEBAF was on such a fast-track schedule, the existing Cornell forming dies were used and their CNC fabrication programs were propagated to vendors for qualification. To minimize the need for laborious mechanical tuning, the frequency was simply rounded to a convenient whole MHz for use in CEBAF. Thus 1497 MHz became the heartbeat of CEBAF. Sundelin, Kneisel, Phillips, and Reece (who joined the Cornell team late in 1983 as a postdoc) began the transition toward moving to Virginia to help build CEBAF. With no facilities yet in Virginia, a formal Cornell-CEBAF collaboration was essential during 1985-1987 in order to support the required R&D and technology transfer/qualification process for potential cavity vendors.[16] The design parameters for the CEBAF cavities are listed in Table 2.

Table 2: CEBAF SRF Cavity Design Parameters

| | |
|---|---|
| Fundamental frequency | 1497.00 MHz |
| Accelerating gradient, $E_{acc}$ | > 5 MV/m |
| Active length | 0.5 m |
| Cell-to-cell coupling | 3.09% |
| Geometry factor | 275 |
| Shunt impedance - $R/Q$ | 960 $\Omega$/m |
| $E_{pk}/E_{acc}$ | 2.56 |
| $Q_{ext}$ input coupler | $6.6 \times 10^6$ ±20% |
| Tuner phase error budget | $10°$ |
| Microphonic phase error budget | $30°$ |
| Lorentz force frequency sensitivity | -2.2 Hz/(Eacc[MV/m])$^2$ |
| Pressure frequency sensitivity | 80–137 Hz/torr |
| Niobium | RRR $\geq$240 |
| HOM $Q_l$ - 1976 MHz mode | 4000 |
| HOM $Q_l$ - 1980 MHz mode | 1800 |
| Beampipe ID | 70.4 mm |
| At $E_{acc}$ = 5 MV/m: | |
| $Q_0$ | $\geq 2.4 \times 10^9$ |
| 2 K dynamic heat load | < 2 W |
| x-plane effective dipole steering | $7.5 \times 10^{-3}$ MeV/c |
| y-plane effective dipole steering | $-1.7 \times 10^{-3}$ MeV/c |
| effective normal quadrupole | $1.2 \times 10^{-3}$ MeV/c/cm |
| effective skew quadrupole | $-1 \times 10^{-3}$ MeV/c/cm |

## V. TECHNICAL CHALLENGES FOR CEBAF SRF SYSTEM

While the design of the basic building block for CEBAF, the 5-cell elliptical Cornell cavity, had passed an essential credibility test in the CESR beamtest, there was yet much work to do to build confidence and resolve associated design challenges with its intended use in CEBAF. Before project construction could be approved, the availability of qualified commercial vendors had to be demonstrated. Within a year, at least three vendors demonstrated successful technology transfer for the fabrication of the cavities for CEBAF. [17-21] For integrated system testing, the cryounit containing one cavity pair and associated helium vessel, tuners, support structure, and RF feedline was the essential package. The hermetic cavity pair concept was adopted and required the development of a cold RF window that becomes integral to the pair assembly. It also required the development of cold RF loads which absorb the HOM power and dissipate it in the 2 K bath.

Until the mid-1980's, the great majority of SRF cavities were fabricated from standard niobium stock that was referred to as "reactor grade," the designation deriving from its use as a structural material in nuclear reactors due to its low neutron capture cross-section and high melting point. Reactor grade niobium has a thermal conductivity of ~8-10 W/mK. As research revealed that peak field limitations for SRF cavities were strongly associated with local material defects, interest grew in using higher thermal conductivity niobium to provide increased thermal stabilization. The more easily measured electrical residual resistivity ratio (*RRR*) is the commonly used experimental surrogate for thermal conductivity, $\lambda$. For the CEBAF cavities, niobium with $\lambda > 60$ W/mK (*RRR* > 240) was specified for the high field regions (cells) while reactor grade material was used for the balance of each cavity.[22]

The cryounit design is depicted in Figure 1.

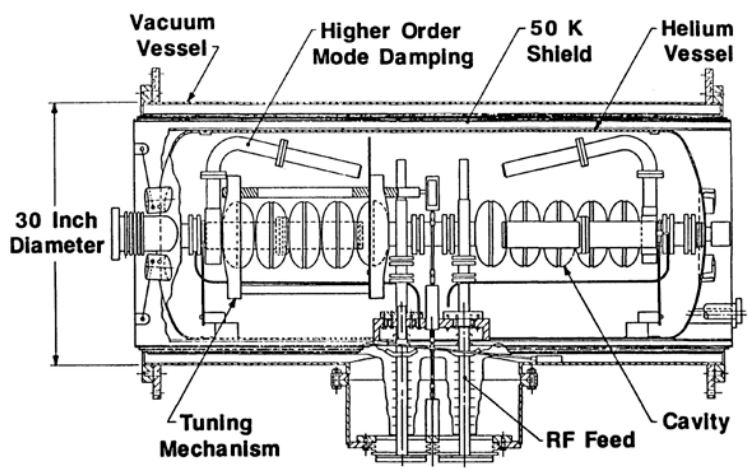

Figure 1. CEBAF cryounit configuration. From [9].

The cavity design was subjected to further characterization regarding its basic beam transport properties,[23] including transverse kicks due to field asymmetries in the input coupler region, [24] dark current,[25] and pressure tuning sensitivities.[26] Later, further analysis confirmed its practical immunity to any multipacting effects.[27]

Initial tests of the cavity pair configuration used Kapton as the cold RF window. Although this design worked, it had significant disadvantages. One is that the double indium layer was subject to cold flow, and, unless adequate force is maintained on the flange joint, the Kapton could pull out of the indium during assembly. (A pressure differential exists across the window during assembly, but not during operation). Other disadvantages are that the Kapton is slightly permeable to gases and is capable of absorbing water vapor; these effects can cause deterioration of the cavity vacuum.

In response to these problems, a hermetic alumina ceramic window was developed. This window had to simultaneously satisfy a large number of requirements: the voltage standing wave ratio must be very low at the operating frequency, it must be reasonably low at other frequencies, the window must be leak tight to both the outside and against through-leaks, the window must not multipactor, the window assembly must tolerate temperature cycling without developing leaks, the window assembly must tolerate slight flexing of its surrounding niobium waveguide without cracking, the window must not be a source of dust or excessive gas, and the power lost by the window must be small compared to the allowed cavity dissipation of 5.5 watts. [21] A window design which satisfied all of these requirements was successfully developed by Larry Phillips. The solution used a custom ground ceramic that was brazed into a thin niobium eyelet, which was subsequently electron beam welded into a bulk niobium frame.[28]

The use of the hermetic pair cavity concept in CEBAF, together with the waveguide higher-order-mode (HOM) couplers of the Cornell cavity created a requirement for RF dissipative loads to be attached directly to each cavity and cooled by the 2 K helium bath. While the waveguide couplers of the Cornell cavity design had successfully coupled out over 280 W of HOM power for dissipation in external water-cooled loads while 22 mA was stored in the CESR beam test, the anticipated HOM power for the CEBAF beam structure was less than 1 mW. Placing the load at 2 K created less cryogenic load than would the static losses of transmission line penetrations. Isidoro Campisi led the development program for the CEBAF HOM loads. Initially, lossy ceramics based on silicon carbide composites were evaluated but were found to have inadequate absorption reproducibility and exhibited undesirable temperature dependent properties below 10 K. An artificial dielectric composed of AlN ceramic sintered with glassy carbon proved very effective. [29, 30]

Because of the very tight stability requirements on energy of the beam delivered by CEBAF, each of the 338 SRF cavities had its individual RF control system with performance requirement of $< 1 \times 10^{-4}$ relative field amplitude stability and 1 degree phase error at 1497 MHz. A unique and economical heterodyned system was developed and implemented.[31]

For optimum power matching to the beam, the design loaded Q value of the CEBAF cavities was 6.6 $\times 10^6$. This implies that the resonance bandwidth of the 1497 MHz fundamental electromagnetic accelerating mode of the cavity was ~ 230 Hz. Establishing and maintaining cavity tune within ±50 Hz requires very careful dimensional control. Allowance for the tuning effects of chemical removal of cavity material, thermal contraction, and pressure sensitivity were well accounted for, so that the 2 K frequency of each cavity was easily targeted within the 200 kHz range of the operational tuner. [32] The mechanical tuner, which grasped the two end cells of a cavity and employed a cold ball screw, was refined to minimize hysteresis and thus extend the anticipated operational life of the rotary feedthrough that connected the warm external stepper motor to the internal gear mechanism.[33]

## VI.   SRF FACILITY SETUP

A significant part of the challenge of establishing CEBAF was the initial complete lack of infrastructure. As mentioned above, the establishment of the collaboration between Cornell University and CEBAF was critical during 1985-1987 when cavity and cryounit prototyping and technology transfer was launched. Niobium superconducting RF cavities for particle accelerators require non-trivial infrastructure investments for fabrication, processing, assembly, and testing. The very desirable low-loss material properties of niobium require a pure, clean, and smooth niobium surface. The basic development work on niobium chemical etching methodologies[34] and the importance of realizing surface cleanliness[35] had been completed, and multiple research labs had experience with testing procedures and the required instrumentation.

Because there was no established industrial supplier of SRF systems or services and CEBAF was to be the first major institution fully dependent on the technology, the establishment of a full-spectrum SRF development and production facility was commissioned.[21] The SRF staff set up operation in the former Space Radiation Effects Laboratory (SREL), which had been operated by NASA from 1960-1980. The presence of this building had been a key to the siting of CEBAF in Newport News. This was a large high-bay building with the remains of a synchrocyclotron vault in its center. [36]

A set of cavity fabrication tools was set up, including sheet metal forming machines, a dedicated CNC mill, and an electron beam welder (EBW). Chemical facilities for cleaning and etching the cavities were established, as was a Class 100 cleanroom where final solvent cleaning of the cavities took place and assembly of the cavity pairs.[22] An existing 20-foot-deep pit, a legacy of the SREL synchrocyclotron, was developed as the location for the Vertical Test Area (VTA). The VTA contains an array of shielded cryostats, plumbed with a closed-cycle liquid helium cryogenic system.[37] All CEBAF cavity development and performance testing has taken place here. To support the production testing of cavity pairs, the first computer-controlled SRF cavity testing system was developed and commissioned.[38] In addition, a cryomodule test facility (CMTF) was established.[39] This facility consists of a shielded cave with cryogenic, RF, and instrumentation necessary to qualify a completed cryomodule as ready for accelerator installation.

## VII.   CEBAF'S IMPLEMENTATION OF SRF

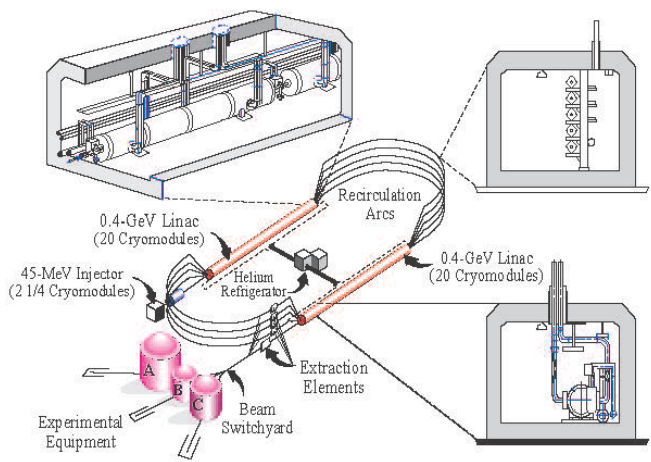

Figure 2. General configuration of CEBAF.

CEBAF's implementation of SRF was built up in the following way. Each hermetic pair of 5-cell niobium SRF cavities was incorporated into one cryounit. Four cryounits were connected with common insulating vacuum and helium plumbing, and bracketed by a pair of cryogenic endcans, the completed assembly, constituting one cryomodule.[40] The endcans provided both the interfaces to supply and return cryogenic flows and the beamline interfaces to warm sections between cryomodules where magnets, beam instrumentation, valves and vacuum pumps were located. There were 20 cryomodules in each of the two linac segments, North and South. There were also two full cryomodules in the Injector, plus an additional "quarter" cryomodule, containing a single cryounit.

The cavities in the linacs are cooled to 2 K by helium gas provided by the Central Helium Liquifier (CHL) and distribution manifolds. The CHL cryogenic load capacity at 2 K was 4.8 kW.

The RF accelerating voltage for each of the 160 cavities in each of the two linacs is set independently. For almost all beam operations, the sum of accelerating voltages from the two linacs is set to match, but is otherwise arbitrary within system capacity. Beam accelerated by 1, 2, 3, 4, or 5 passes through the linacs can be extracted and simultaneously routed to experimental Halls A, B, or C. The amount of current delivered to each experimental hall is also set independently. Over 2,030 magnets were used to direct and focus the electron beam in CEBAF. Each of these is controlled by a precision power supply.

The fruit of this complexity is a very flexible and efficient tool for nuclear physics research.[41, 42]

## VIII. SRF PRODUCTION FOR CEBAF CONSTRUCTION

### 1. Cavity fabrication

The procurement contract for 360 5-cell niobium SRF cavities for CEBAF was awarded to Interatom GmbH.[43] The niobium material stock was independently purchased by CEBAF and provided to the fabricator. The high purity material (residual resistance ratio (RRR) >250) used for the cavity cells came from three different suppliers. W. C. Heraeus, Teledyne, and Fansteel Corp. provided 3%, 20%, and 77% of this stock, respectively.[32] The cavity contract was for fabrication only. The only active chemistry was a light acid etch on the approximately 40 niobium parts that went into each cavity just prior to EBW steps. All chemistry on the finished cavities was accomplished after delivery to CEBAF.

The bulk etch of the cavities was accomplished by immersion in an acid bath consisting of equal parts of hydrofluoric (48%), nitric (69%), and phosphoric (86%) acids, also known as Buffered Chemical Polish (BCP) 1:1:1. The pre-chemistry consisted of two steps of 2.5 minutes each, with intermediate cavity rinsing to avoid overheating the acid bath. Each step removed approximately 25 µm from both inner and outer surfaces. After final frequency tuning, a final chemistry treatment of 1 minute lowered the cavity frequency by ~70 kHz and yielded the finished surface.[32]

Because the RF waveguide fundamental power coupler (FPC) was also the principal mechanical support of the cavity pair, considerable attention was given to final dimensions of the sheet-metal fabricated cavities. In many cases, custom machining of the in-house fabricated interface parts was required.[44-46]

While the production fabrication of the CEBAF superconducting cavities was accomplished by a vendor, the niobium components which attach to the cavities were fabricated predominately using CEBAF in-house facilities. One ceramic RF window, brazed and EB welded into a niobium frame was required for each cavity. One all-niobium FPC with 38 tapped holes in the main interface flange was required for each cavity. Two waveguide elbows that couple beam-stimulated higher order mode power out to the HOM loads were required for each cavity. These also were of all-niobium composition. Finally, each cavity pair required a niobium beampipe between the cavities. All of these >1500 niobium pair parts were fabricated at Jefferson Lab.[47] The HOM loads described earlier were also all fabricated at JLab. [29]

## 2. Cavity testing

Between May 1990 and September 1993 over 275 RF tests were made on cavity pairs at 2.0 K.[48] From these tests, 15 pairs were rejected due to poor RF performance and were subsequently successfully reworked. The hardware configuration for the original CEBAF cryomodules places the entire cavity pair assembly within a large helium vessel. Each cryounit then has 22 indium seal gaskets between superfluid 2 K liquid helium and beamline vacuum. Confidence in the flange surface preparation to assure leak-tight joints had to be developed. A new He desorption leak detection technique with a sensitivity of $10^{-15}$ std-$cm^3$-sec was routinely used to verify indium seal integrity.[49] An analysis of the production issues encountered and addressed during CEBAF construction has been provided by Mammosser.[50] These were principally associated with vacuum leaks and particulate contamination control.

The tested performance of the CEBAF cavity pairs greatly exceeded the requirements but with a very wide distribution.[48] Figure 3 shows representative performance limitations encountered during the production vertical cavity pair testing. Either a thermal-magnetic quench provided a hard limit on the stored energy, or electron loading degraded the cavity $Q$ intolerably. Subsequent investigations many years later associated the observed characteristic quench distribution of these cavities (shown in Figure 4.) with common flaws in the cell equator welds linked to residual contamination during fabrication. The electron loading was credited to residual surface contamination after the chemical processing and assembly at CEBAF. The distribution of the performance at onset of observed external x-radiation was also noted. (Figure 5.)

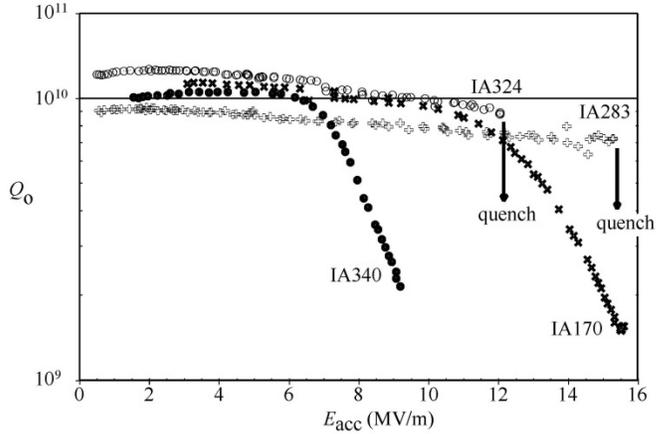

Figure 3. Representative performance of CEBAF cavities in 2 K acceptance tests. From [51].

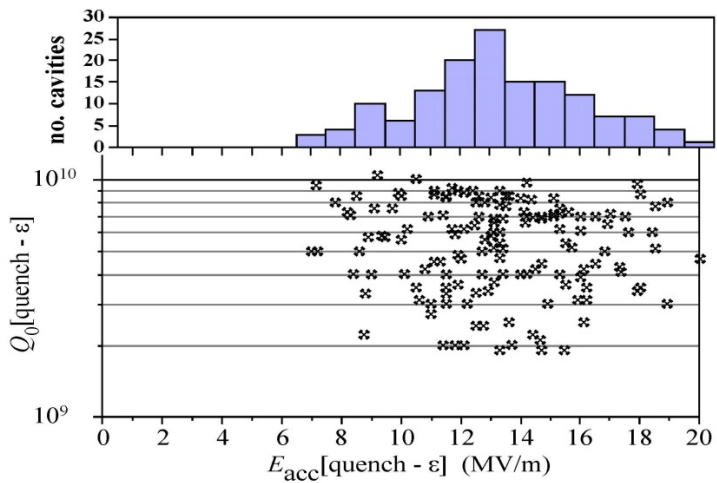

Figure 4. Cavity performance just below quench during pair testing. From [48].

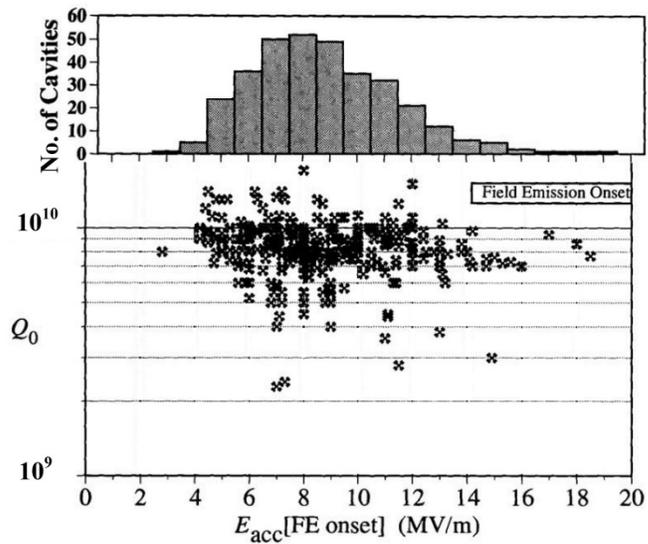

Figure 5. Cavity performance at onset of detected x-rays during pair testing. From [48].

### 3. Cryomodule assembly

In order to preserve the very tight energy and beam spot tolerances required of the CEBAF cryomodules, it is necessary that the cavity centerlines be aligned to within 2 mrad (rms) of the beamline. Procedures and tooling were developed to meet this requirement and preserve it with installation and cooldown of the cryomodules.[52]

Assembly of cryomodules began in 1990. By mid-1992, 14 cryomodules had been completed and installed in the North linac. Four cryounit assembly lines ran in parallel, each producing one cryounit every two weeks. Finished cryounits fed two cryomodule assembly lines, each producing one cryomodule per month. This net production rate of two cryomodules per month was established in early 1992 and maintained through completion of the project.[53] The thermal performance of the cryomodule design was carefully characterized and found to consistently beat the budgeted 15 W static heat leak. [54] Due to limited cryogenic capacity in the Test Lab, very significant remanent magnetic field in the cryomodule test cave left behind from its previous use as a synchrocyclotron vault, and tight schedule pressure, all but the first few cryomodules were taken directly from assembly and installed in the linac tunnel. Their basic performance commissioning tests were performed in their operational locations.[55]

### 4. Cryomodule commissioning

The overall CEBAF installation and commissioning exercise was highly orchestrated, with several activities running in parallel.[53, 56] The short injector linac was commissioned first and used for an experimental check of multipass beam breakup instability and energy recovery by implementing a local beam recirculation line.[57] An energy vernier system was developed and tested in the north linac for stabilizing the relative energy spread to better than $2.5 \times 10^{-5}$.[58] Also, high current tests were implemented to check that current modulation effects did not result in beam energy variability.[59] Cryomodule commissioning tests were all coordinated by Mike Drury. The bulk of commissioning of the SRF cryomodules in both the north and south linacs was completed in 1992-3.[50, 60] By this point all of the hardware, instrumentation, and controls details had been thoroughly debugged. The cryomodule performance well exceeded the CEBAF design requirements, despite the fact that there were a few non-functioning cavities.

It is usually the case for particle accelerators constructed as research tools that the user community is happy to use a facility that meets the required project specifications, but is equally interested in exploiting all capability that happens to be available beyond the formal requirements. This was quite true in CEBAF's case. There was keen interest to characterize and understand the performance limitations of the SRF cavities, to learn how to fully exploit what was available, and to set in place mechanisms for continuous improvement. The central performance-limiting phenomenon for the installed SRF cavities in CEBAF was electron field emission internal to the cavities. In addition, some of the cavities exhibited uncharacteristically low Q's, resulting in more than double the expected cryogenic heat load for those cavities. An analysis of the performance of cryomodules in the south linac pointed to multiple contributing effects: slow rate of cooldown for some locations resulting in the so-called "Q-disease" due to the ready formation of lossy niobium hydrides in the region of 100 K, excessive magnetic field environment at a cavity location, and excessive RF losses in the cold RF window assembly or copper plated waveguide in the 2 K-to-room temperature transition.[61]

## IX. PERFORMANCE OF CEBAF SRF CAVITIES FOR 4 GEV

As mentioned above, the vendor-supplied CEBAF cavities significantly exceeded requirements. Because of the very high priority for the new laboratory to not disappoint the nuclear physics users, the baseline project performance specifications were set deliberately quite conservative: accelerating gradient of 5 MV/m and $Q_0$ of only $2.4 \times 10^9$. Just what the actual installed performance capacity turned out to be was of keen interest and was reported to the Particle Accelerator Conference in 1995 [51], summarized here.

In order to use the cryogenic capacity most efficiently and to minimize steady-state x-radiation fields, the *usable* gradients were initially constrained to fields which produce less than 1 watt of cryogenic load dissipation attributed to field emission loading. This constraint applied to the cavity pair qualification tests resulted in the distribution shown in Figure 6. This distribution clearly falls between the field emission onset and quench condition distributions of Figures 5 and 4.

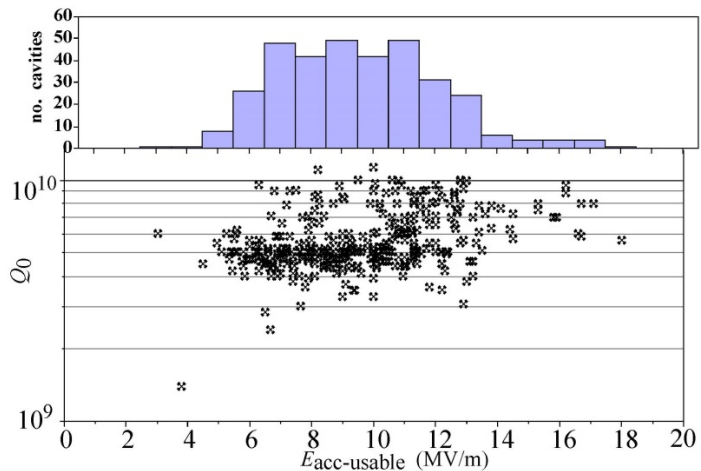

Figure 6. CEBAF cavity usable performance per vertical acceptance tests. From [51].

As the cryomodules were commissioned, a maximum allowed operating gradient was established, bounded by additional constraints: quench field minus 10%, 1 rad/hr measured outside of the cryostat in the tunnel, waveguide vacuum excursions, and window arcs that prevented 1 hour sustained operation. In addition, available RF power in the accelerator limited the maximum no-current gradient to about 14 MV/m. The resulting distribution of maximum *usable* gradients determined from cryomodule commissioning is shown in Figure 7.

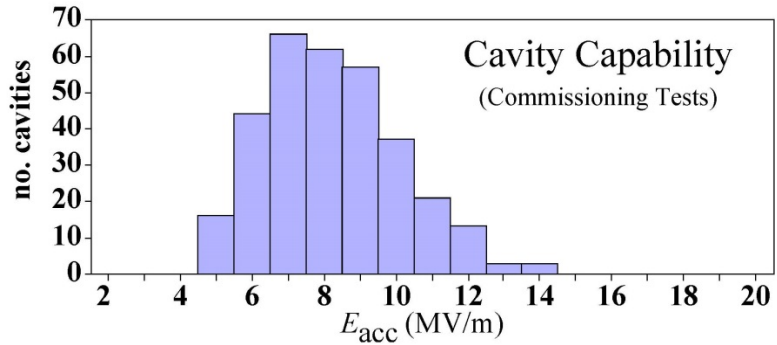

Figure 7. CEBAF cavity usable performance per tunnel commissioning tests. From [51].

The CEBAF ceramic RF window is mounted directly to the cavity waveguide fundamental power coupler. The alumina ceramic is located only 79 mm off of the beamline. In this location, the 2 K window is subject to charging via electron and x-ray flux. During sustained cryomodule operation after their initial commissioning, quite a few cavities exhibited "arcing" in the region of the cold window at a rate which was otherwise unacceptable for operations—as high as 45 times per day. An additional constraint of < 2 arcs/day was then added to the criteria for *usable* maximum gradient for each cavity. This operational derating was promptly necessary for 13% of the cavities installed in CEBAF.

Based on the initial low-power operating experience in 1995, the distribution of types of cavity gradient limitation was as shown in Figure 8. Clearly, field emission loading and the periodic arcing phenomenon represented the most significant gradient performance limitations for CEBAF.

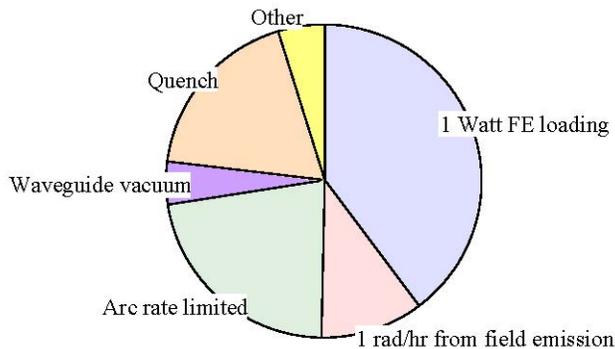

Figure 8. Distribution of CEBAF cavity gradient limitations, 1995. From [51].

This "arcing" became recognized as clearly correlated with the presence of nearby field emission—either in the arcing cavity or its neighbor. When such arcing does occur, its frequency is strongly dependent on cavity gradient. Note that other cavities functioned stably, without arcing, above 9 MV/m. The analysis and operational adaptation to this phenomenon is the subject of the next section.

The initial overall view of the capability of the installed CEBAF cryomodules during limited low-power operation in 1995 and projected into full 5 kW/cavity operation was captured in Figure 9.

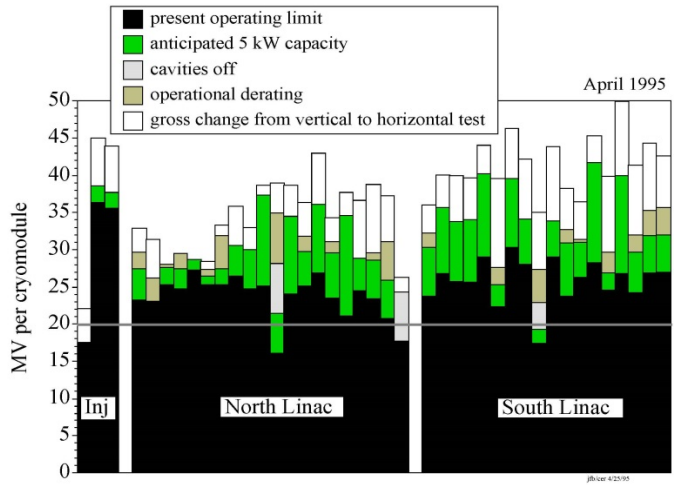

Figure 9. Operational projection of CEBAF cryomodule capacity in 1995. From [51].

The integrated system performance was shown to be excellent as CEBAF was commissioned for beam operations.[62] Automated routines were developed for tuning and phasing all of the cavities in each linac. The recirculation arcs 1 and 2 were used as sensitive spectrometers to calibrate the effective accelerating gradient of each cavity individually to better than 2% and then to stabilize the delivered beam energy to better than $5 \times 10^{-5}$. CEBAF delivered beam to its first nuclear physics experiment in Hall C in November 1995 and continued to deliver up to 300 kW of beam power at various energies. Measured beam emittance was 0.2 nanometer-radians, better than the design specification.[63]

## X. ARC TRIP PHENOMENON CHARACTERIZATION

Initial enthusiasm that the as-installed CEBAF SRF cryomodules would perform easily at more than 50% above the project requirements dissipated rather quickly as operational experience began to accumulate. Stable RF system performance was frequently interrupted by trips of the input RF window protection arc detector and input RF waveguide vacuum interlocks. The interlocks are required to quickly disable the supply of RF power in the event that either a discharge or arc event and/or the waveguide vacuum exceed some threshold pressure. Sustained discharges can critically damage the windows and compromise the beamline vacuum. Input RF power coupling typically requires deliberate technological care.[28]

Investigations were launched to characterize the phenomenon, and subsequently strategies were developed to maximize the useful accelerating voltage of CEBAF. Empirically, it was found that reducing the operating gradient of an offending cavity reduced the arc occurrence rate. Since CEBAF had generous voltage margin for its required operation at 4 GeV, offending cavities were derated and commissioning operations proceeded.

Offline investigations included (a) examination of the time structure of the RF power during such events[64], (b) spectroscopic analysis of the arcs which clearly associated light from periodic arcing with the ceramic material in the cold RF window[65, 66], (c) a detailed distinction of five different types of concerning events[67], (d) careful analysis of local charging currents[68], and (e) modeling of field emitted electron trajectories that can reach the cold ceramic window.[69] In all cases, discharges occur only in the presence of field emission and occurred with roughly a constant charge accumulated on the window. The bulk resistivity of the alumina ceramic at 2 K is extremely high, so the natural current bleed-off is extremely slow.

Operationally, Jay Benesch took up the challenge of developing gradient-dependent models for the periodicity of arc events as a function of sustained operating field level for all of the affected cavities in CEBAF.[70] Use of a functional parametrization akin to Fowler-Nordheim theory for the field dependence of emission current as the driving term for charge accumulation on the cold windows proved quite useful. Acquiring statistically significant data for arc rate as a function of cavity voltage for each cavity, $R_i(V_i)$, was difficult. The most relevant data correspond with $V_i$ that yield $R_i$ ranging from 2 hr to 5 days. This required extended periods of stable operation and that the bulk of the data accumulation be necessarily parasitic to accelerator operations. As such data was acquired, it was parametrized assuming

$$R_i(V_i) = e^{a_i + b_i V_i}$$

The integrated system arc trip rate for a given linac energy gain setup ($\Delta E = e \sum_i V_i$) is obtained when $t_R = \partial R / \partial V$ is uniformly constrained. This parametrization was then incorporated into the machine setup routine by Larry Doolittle and established the route to maximizing the available CEBAF energy as a function of maximum tolerable trip rate for the nuclear physics experiments.[71]

Analysis of data in 1998 produced the distribution of individual cavity accelerating gradients which produce periodic arc trips at 4-day intervals shown in Figure 10. This integrated to a machine trip rate of 3.25 faults/hr, quite acceptable for the nuclear physics experiments and yielded a very solid ~5.2 GeV accelerator system.

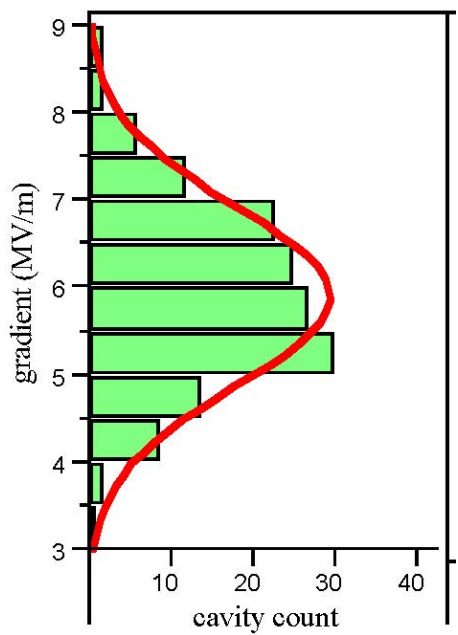

Figure 10. Modeled gradients for arc fault intervals of four days. From [70].

Benesch continued to track the evolution of the cavity-specific trip rate models over the subsequent two decades to quantify operational benefits of RF cavity processing efforts such as helium processing and spontaneous changes to the field emission source terms in cavities, typically associated with migration of particulate contamination.[72-77] This has provided vital feedback on machine reliability and capacity.

Minimization of the operational downtime associated with the arc interlock trips also involved analysis of constraints on rapid reset. The energy acceptance of the recirculation arcs is significantly less than the energy gain contributed by a single cavity, so each arc trip requires interruption of beam delivery as well as turn-off of the offending cavity RF power. A physical effect of each arc is liberation of adsorbed gas from the adjoining waveguide. This in turn generates a consequential fault in the waveguide vacuum interlock. Analysis of the vacuum recovery time has indicated that automatic reset after as little as 0.5 second may be possible.[78]

It is certainly worth mentioning that this window arcing phenomenon is a performance feature unique to CEBAF. It is an unexpected consequence of the early decision to build CEBAF from hermetic cavity pairs, with cold ceramic windows only 7.9 cm off of the beamline, together with the push to operate cavities above the 5 MV/m design gradient in the presence of sustained field emission. Subsequent designs effectively eliminated all such periodic arcing by displacing the cold ceramic window farther from the beamline and avoiding line-of-sight exposure to any free electrons in the cavity.[79]

## XI. CEBAF INITIAL OPERATION CAPACITY

### 1. First full physics operations

The process of commissioning the full capability of CEBAF continued in 1995 and 1996 in parallel with first runs for physics.[63, 80] Establishing accurate as well as flexible setup of over 2200 magnets, 300 SRF cavities, 40,000 control points, and 120,000 database records required development of sets of automated calibration and setup routines. Some early operational highlights included: delivering 25 µA CW beam at five discrete energies in an eight hour period; delivering beam from the polarized source to an experimental hall; running at greater than 1 GeV for a single pass; delivering 4.0 GeV, 80 µA CW beam to an experimental hall; delivering RF separated beam to Hall A and Hall C simultaneously; and sending three separate 60 µA CW beams to a dump.

The as-built CEBAF linacs were capable of solidly supporting 5.2 GeV, 5-pass operation. In order to save on electrical power costs during commissioning, the available klystron power was reduced. The initial voltage capability of the SRF cryomodules is depicted in Figure 11.

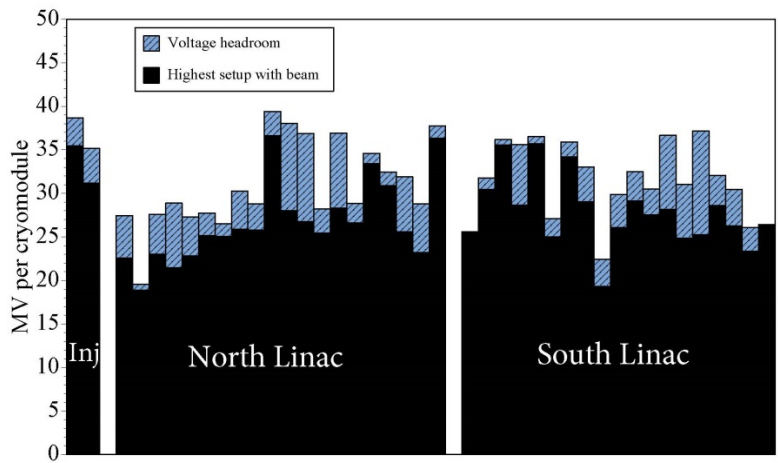

Figure 11. CEBAF available SRF cryomodule voltage in 1996. From [63].

An important aspect of exploiting SRF cavity systems for particle accelerators is the RF control system which must regulate frequency, amplitude, and phase stability. CW accelerators aim to provide "always on" stabilized accelerating voltage. The CEBAF low-level RF controls (LLRF) were successfully modeled and tested to thoroughly understand challenges to regulation as a function of beamloading, both CW beam and pulsed beam.[59, 81] One very attractive feature of SRF CW accelerators is the potential for high power efficiency with the supplied RF. A very large fraction of the supplied RF power is directly converted into beam power, with very little dissipation in the resonator. This high beamloading circumstance places extra demand on the LLRF when delivered beam energy precision is paramount. During the commissioning run period, the reduced klystron power availability translated into a maximum beam current envelope for different number of recirculation passes and final beam energy. Figure 12 shows the operational envelope for CEBAF in 1996.

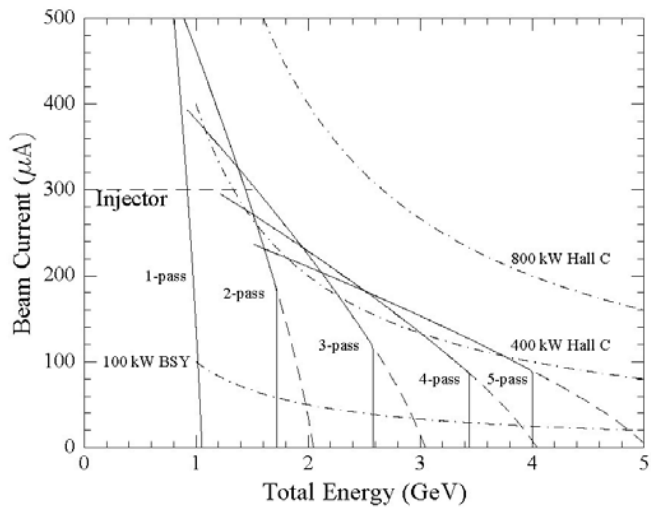

Figure 12. 1996 operational envelope of CEBAF at reduced klystron power. From [81].

## 2. *In situ cryomodule improvements*

Because the principal performance constraint on the SRF systems in CEBAF was derived from the occurrence of field emission from the cavity surfaces—x-ray production, charging and arcing at the cold ceramic RF window, and anomalous 2 K heat load—there was keen interest to find *in situ* processing techniques which could reduce these effects. While the processing of SRF cavities with low pressures of admitted helium had been a technique used in test settings for many years,[82, 83] there seemed to be no experience with its use on cavities installed on an accelerator beamline. The first attempt in CEBAF was its application to a very poor performing cryomodule that was soon to be pulled out of the machine for rework, NL03. This "helium processing" operation was so successful at improving performance that the cryomodule was left in the machine and operated usefully for another decade.

Subsequently, an increasingly structured approach was developed for systematically applying helium processing *in situ* to the cavities in CEBAF.[84, 85] By August 1998, the process had been applied to 224 cavities in 29 of the 41 installed cryomodules. Significant progress was made in both the reduction of observed x-radiation from field emission and also the arcing phenomenon. Figure 13 shows the distribution of maximum gradient for this set of processed cavities according to their performance-limiting constraint.

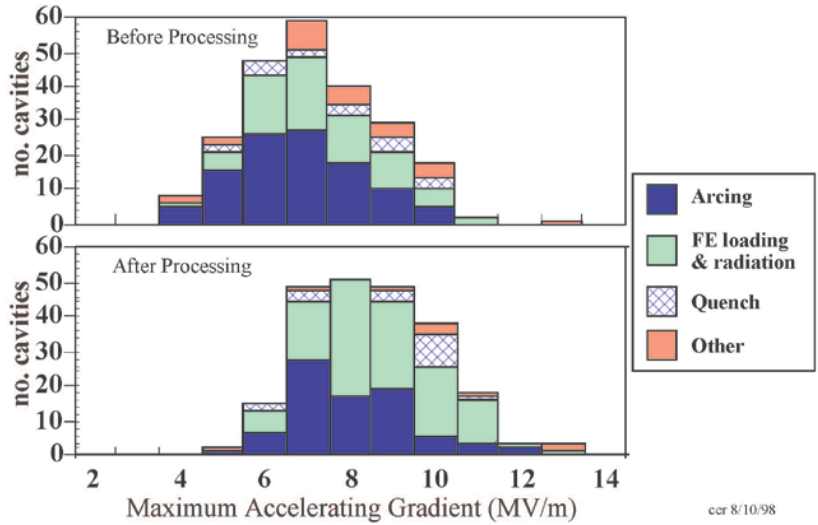

Figure 13. Cavity performance improvement with helium processing 1996-1998. From [85].

This helium processing and requalification of cavities increased the installed CEBAF voltage by 155 MV, corresponding to an added 775 MeV for 5-pass beam. To help gain systematic insight into the ensemble of independent cavity systems which may be constrained by an array of factors, Jean Delayen developed an analysis that helps disentangle the various factors and prioritize remediation work by modeling anticipated benefits of progress against one or another individual constraint.[86] In addition, the operational setup of the linac voltage distribution was managed by minimization of arc trips, anticipation of beamloading compared to available klystron power, and cryogenic load. Each of these is a function of individual cavity gradient. Larry Doolittle's linac energy management setup program "lem++" was the first to tackle such a multi-parameter setup challenge in an SRF linac.[71]

### 3. Pushing the performance envelope

In September 1997, a test was performed on the CEBAF accelerator to demonstrate its full design capability of 200 μA CW beam at 4.0 GeV. This beam was delivered to the Hall C dump with minimal difficulties and with stability characteristics consistent with normal operations. Since no physics target was available that could handle 200 μA, the test was performed without a target in place. Rastered beam went directly to the dump. One clear indicator of the robustness of the integrated accelerator system was the ability to place 200 μA on the injector Faraday cup, retract that cup, and immediately transport lossless, full-power beam to the Hall C dump and maintain this current for at least tens of minutes with all orbit locks off. The duration of the full test was 10 hours. As illustrated in Figure 14, the test demonstrated the economy of exploiting SRF technology. A typical cavity operating at 6.5 MV/m fully matched the input 3.8 kW RF to the beam. During this run the measured peak-to-peak fractional energy variation was $4 \times 10^{-5}$.

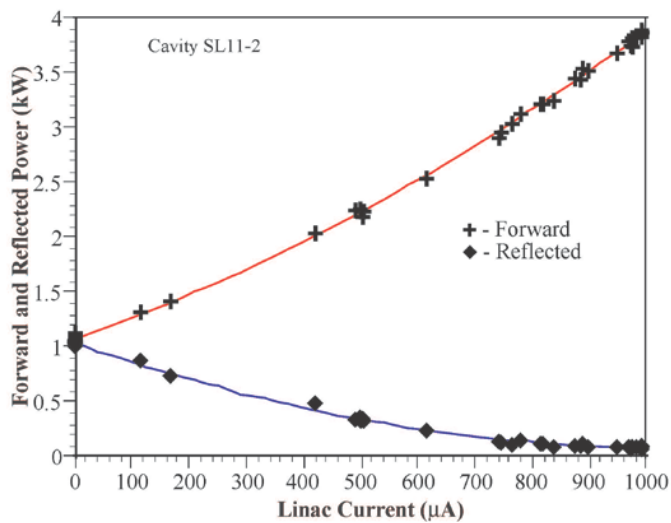

Figure 14. Matched RF load at full beam current. From [85].

It is interesting to note that while a good bit of attention was going toward increasing the manageable beam power in CEBAF, simultaneously skill was being developed to deliver very low CW currents to Hall B, as low as 100 pA—which corresponds to an average of 1.25 electrons per 499 MHz RF bucket.

During the January 1999 machine maintenance down an additional eight cryomodules received *in situ* helium processing that yielded an additional 52 MV benefit. Progress in the evolution of demonstrated maximum voltages from the CEBAF cryomodules by 1999 is illustrated by Figure 15.

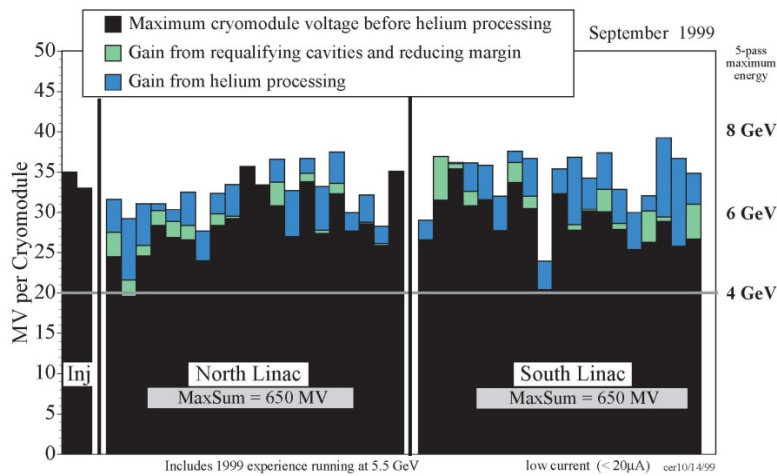

Figure 15. Maximum SRF voltage per cryomodule in CEBAF in 1998. From [87].

During fiscal year 1999, CEBAF provided 5360 physics beam hours with an average active-use multiplicity of 2.6. This included delivery of beams throughout the 0.8–5.5 GeV range. Use of polarized electrons had become standard. As accumulated knowledge of the individual cavities improved, the optimization algorithms were able to decrease the number of arc trips per day during the 5.5 GeV physics runs from ~220 to ~60. By 2001 almost all of the cryomodules had received helium processing and reasonably thorough characterization of the gradient dependence of arcing had been established for 211 of the 315 installed linac cavities. Figure 16 shows the distribution of the gradients producing a periodic arc every 8 hrs. The performance was well above the original 5 MV/m requirement and predominately limited by the effects of field emission derived, presumably, from residual internal particulate contamination from the initial linac construction.

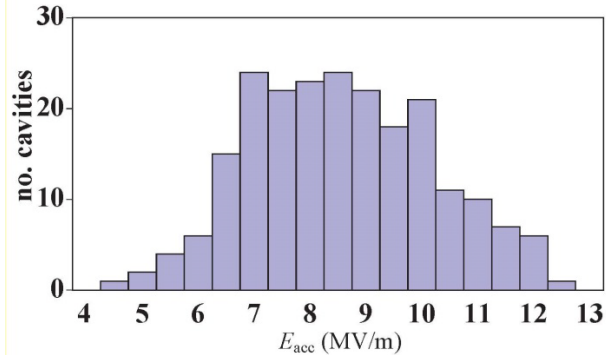

Figure 16. Distribution of cavity gradients producing 3 arcs per day. From [88].

The progression of successful demonstrations of the beam energy and power envelope for CEBAF is depicted in Figure 17.

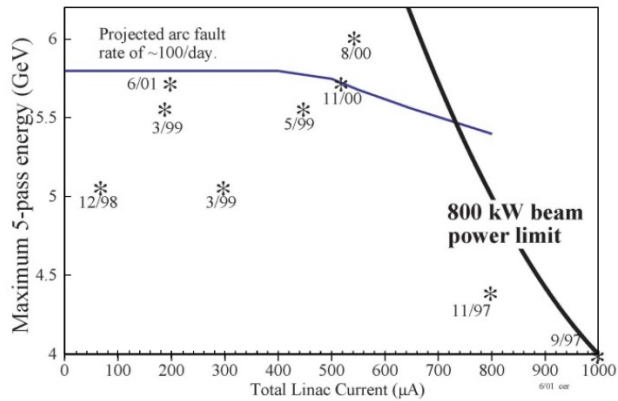

Figure 17. CEBAF operating highpoints prior to 2002. From [88].

In 2002, CEBAF was operating extremely well for multiple physics users and was eventually pushed to deliver 5.85 GeV physics beam. A test run achieved 6.0 GeV, but with stability inadequate for physics use.[89]

## XII. IMPACT AND RECOVERY FROM HURRICANE ISABEL

In September 2003 Hurricane Isabel passed over the Hampton Roads, Virginia area. Jefferson Lab was subsequently left without electrical power for four days. This was long enough to lose insulating vacuum in the CEBAF cryomodules and cryogenic systems resulting in the whole accelerator warming up to ambient conditions and the total loss of the liquid helium inventory.[90, 91] This thermal cycle stressed many of the cryomodule components causing several cavities to become inoperable due to helium-to-vacuum leaks. At the same time the thermal cycle released years of adsorbed gas from the cold surfaces. Over the next few weeks this gas was pumped away, the insulating vacuum was restored and the machine was cooled back down and recommissioned. In a testament to the robustness of SRF technology, only a small loss in energy capability was apparent, although individual cavities had quite different field-emission characteristics compared to before the event.

Most of the cryomodules had never been warmed up since initial cooldown in 1991-1993. The 2003 warm-up released gas that had cryosorbed on the insulating, waveguide, and beamline vacuum surfaces during the intervening decade. Within a month, all cryomodule vacuum spaces were reestablished, and all but one cryomodule was cooled down and returned to service. New helium-to-waveguide vacuum leaks appeared in the north linac zone 5 cryomodule, requiring four cavities there to be bypassed. One other cavity developed an open field-probe cable and could not be recovered. Otherwise RF control was reestablished for all cavities by the end of October. Performance stability as a function of cavity gradient was re-explored and new arc rate vs. gradient models were constructed for all cavities.

Before the Isabel shutdown, CEBAF had been running at 5.5 GeV with ~5 RF trips per hour. This rate was approximately doubled post-Isabel. In order to satisfy a committed physics run at 5.75 GeV, a concerted push was made to refine and optimize the setup and maximize the up status of all RF systems, with the result that with all usable cavities on, the trip rate was as low as 10 trips/hour at this energy in January 2005. It was very interesting to note the effective randomization of arc-rate limitations of the cavities across thermal cycles to room temperature. The model calculations of individual cavity gradients corresponding to 1 trip per day before and after the hurricane interruption are plotted in Figure 18. One may speculate that the mix of improvements and degradations is the fruit of a combination of desorption of field emission enhancing adsorbed gas on emitters and possible migration of particulate emitters during beamline servicing and re-evacuation.

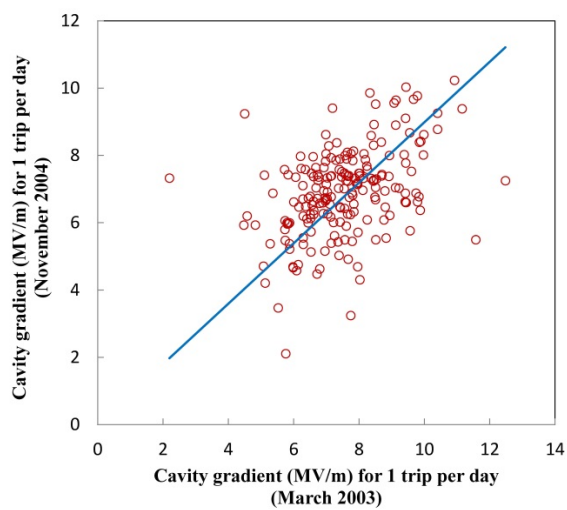

Figure 18. Comparison of fixed arc-rate gradient conditions before and after the hurricane-induced thermal cycle and recovery. From [90].

## XIII. CEBAF PERFORMANCE CIRCA 2004

By mid-2004 CEBAF was solidly back in operation supporting an active physics program. The versatility of CEBAF was quite remarkable. This can be illustrated by pointing to the delivery of three different "world's-best-quality" beams to three different experimental halls – simultaneously.[92] A high level of sophistication had been established in the CEBAF injector such that the requirements listed in Table 3 were fully met or exceeded simultaneously to three very different experimental programs.

Table 3: CEBAF Experiment Beam Requirements 2004

| Beam Parameters | Hall A | Hall B | Hall C |
|---|---|---|---|
| Current | 100 µA | Few nA | 40 µA |
| Charge/bunch | 0.2 pC | $2 \ 10^{-17}$ C | 1.3 pC |
| Energy | 1-5 GeV | 1-5 GeV | 1-5 GeV |
| $\Delta E/E$ | $2.5 \times 10^{-5}$ | $5 < 10^{-4}$ | $5.0 \times 10^{-5}$ |
| Size at target | >100 µm <1000 µm | 100 µm | <200 µm |
| Divergence | -- | 100 µrad | 100 µrad |
| Fractional Beam Halo | 100 Hz/µA @ $R = 3$ mm | $<10^{-4}$ @ >0.5 mm | $<10^{-6}$ @ $R > 3$ mm |
| Polarization | -- | -- | >70% |

Earlier, the whole of CEBAF was used to implement a 1 GeV energy recovery experiment.[93] An extra $\lambda_{RF}/2$ chicane was added in the arc 2, so that the beam was 180° out of phase with the accelerating RF on its re-injection into the north linac and the beam power was then transferred back into the cavity RF fields. A new temporary beam dump was set up at the end of the south linac to dump the beam at the original injection energy. The experiment was run with two different injection energies: 55 MeV and 20 MeV. A full 80 µA CW operation was achieved with the 55 MeV injection. The prime attractive property of SRF energy recovering linacs is illustrated by the data from this test depicted in Figure 19. During CW beam operation, the required forward power driving each cavity is essentially the same as no-beam operation. The short pulse at the beginning and end of the energy recovered data are due to the 4 µsec transit time of the machine.

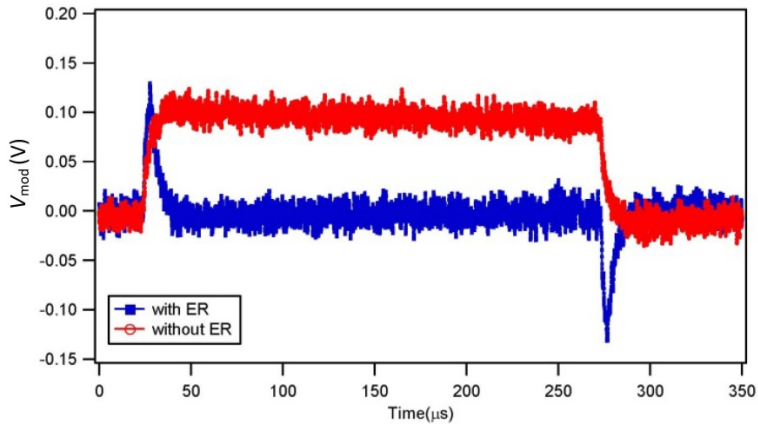

Figure 19. Forward power drive signal for an SRF cavity with 250 μsec-duration tune beam with single beam (red) and a two-beam, energy recovering setup (blue). From [93].

## XIV. SUPPORTING SRF INFRASTRUCTURE IMPROVEMENTS

In order to continue to improve deployable SRF technology for CEBAF and other DOE laboratory projects, there have been a few cycles of investment in infrastructure upgrades at Jefferson Lab. With the very clear interest in reducing the impact of field emission on operational SRF accelerators, improvements were made to the final cleaning processes for the niobium cavities. This included upgrades to the ultra-pure water (UPW) system and commissioning of a high pressure rinse (HPR) spray system with which to scrub particulates from the sensitive surfaces just prior to assembly within the cleanroom.[94-96] The original CEBAF cavity final cleaning step involved simple rinsing with clean methanol. As analytical techniques were brought to bear on quantifying residual particulates, it became clear that water quality, ambient environment, and even drying orientation had significant influence on ultimate cavity performance.[97, 98] Use of HPR with UPW was introduced during the latter half of CEBAF construction, and use of the automated tool became a standard part of the JLab SRF cavity preparation protocol after 2003.

Safety and process control was also improved by the addition of a PLC-controlled closed chemistry cabinet in which the acid etching of niobium cavities was applied. This new cabinet was located directly in the Class 100 cleanroom, replacing the dipping of cavities in an open vat of acid in a non-clean environment.

As achievable accelerating gradient continued to increase, it became apparent that, at least for the standard fine-grain niobium material, microscopic surface smoothness became an important influence on ultimate performance. Electropolishing of niobium cavities, following the pioneering work at KEK[99], became very attractive. As part of the JLab contribution to the SNS linac construction, a production SRF cavity electropolishing system was set up and commissioned in 2003.[100]  Several incremental improvements were made to the system over the subsequent decade as it saw use for both the ILC R&D effort and the later CEBAF Upgrade.

The JLab cryomodule test facility (CMTF) was originally implemented in the late eighties for testing of a small fraction of the cryomodules during the production run for CEBAF.[39] The original system was built using a dedicated wiring scheme and a pair of 2 kW, 1497 MHz RF sources. This dedicated system made it difficult to test cryomodules and other RF structures of non-standard configuration. Additionally, due to the previously installed synchrocyclotron, there were remanent magnetic fields in excess of 6 Gauss within the test cave, severely limiting the capability of the facility when measuring the quality factor of superconducting cavities. Testing of the Spallation Neutron Source cryomodules as well as future upgrades to the CEBAF accelerator necessitated a complete renovation of this facility.[101] The company Vitatech Electromagnetics, LLC was hired to install layers of magnetic shielding inside the test cave. The ambient magnetic fields were successfully reduced to less than 200 mG over the full cryomodule test volume. A new set of flexible controls hardware and software was commissioned and used for testing the SNS cryomodules and all subsequent CEBAF cryomodule work.

As the technologies that support web-based information management (IM) blossomed, JLab SRF sought to exploit them to enhance labor-efficient workflow, quality assurance, and continuous improvement as a learning organization. With very little institutional support, the Accelerator Systems Department launched a small effort to create an IM system with which to orchestrate the complex set of procedures and technical data collected during the development and production of SRF accelerators. This infrastructure development proved as vital as any piece of hardware. The "Pansophy" system, as it came to be called, provided a convenient user interface with which to both structure and use a robust underlying database.[102-104] The system provides the tools for electronic "travelers" to be created directly by engineers and scientists. These version-controlled travelers are both "prescriptive" and "descriptive", in that they specify actions and procedures, and also collect data and part-specific comments from each implementation of a task. Since its initial launch Pansophy has continued to evolve to meet new project needs. All of the data collected by Pansophy remain accessible for after-the-fact mining.

## XV.  C50 CRYOMODULES FOR 6 GEV

As described above, after Hurricane Isabel, the energy reach of CEBAF was limited primarily by the field emission-induced charging and arcing on the cold ceramic RF windows. A decade after the completion of CEBAF cryomodules a project was launched to refurbish the ten worst-performing cryomodules using the then-best available processing techniques and to make modifications which would eliminate occurrence of the periodic arcing phenomenon from the reworked cryomodules. The klystrons and LLRF installed in CEBAF could support operation of the 0.5 m, 5-cell cavities to about 13 MV/m with the required beamloading for physics. The performance specification for the reworked cavities was then set to 12.5 MV/m, which results in an eight-cavity cryomodule providing 50 MeV/pass.

The cryomodules were thus labeled "C50s" to distinguish them from the original cryomodules which typically provided about 30 MeV/pass.[105]

The project called for reworking three cryomodules per year in order to establish a solid 6 GeV physics capability that would in turn serve as the base for a later upgrade to 12 GeV.[106] The project ran from 2006 through 2009. During the refurbishment process, each cryomodule was disassembled and its cavities removed and refreshed with improved processing techniques using vacuum furnace degassing and the new chemistry and HPR facilities. The fundamental power coupler waveguide section was redesigned to displace the cold ceramic window outward and to include an intervening "dogleg" in the waveguide between cavity and window. (See Fig. 20.) The original polyethylene warm RF window was replaced with an improved ceramic model. Improvements were made to the mechanical tuners to reduce backlash, and components that were subject to mechanical wear or radiation damage over the years were replaced as well.

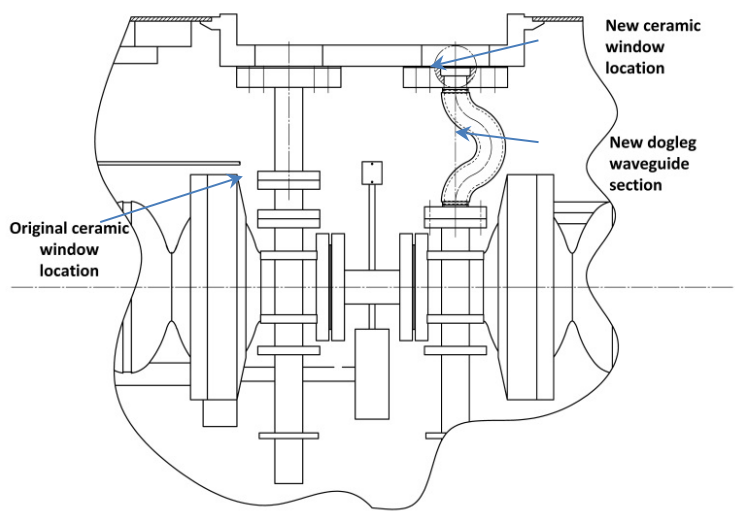

Figure 20. RF window relocation and dogleg waveguide in C50 cryomodules.

The project was quite successful in dramatically reducing the presence of field emission loading in the reworked cavities and in almost completely eliminating the occurrence of the arcing phenomenon upon reinstallation in CEBAF. The cavities were then more typically field performance limited by quenching associated with original fabrication flaws in the equator welds, but still exceeded the 12.5 MV/m goal 70% of the time.

Figure 21 shows the evolution of the CEBAF maximum 5-pass beam energy from 1995 through 2011, including the recovery of full 6 GeV capability from this C50 project.

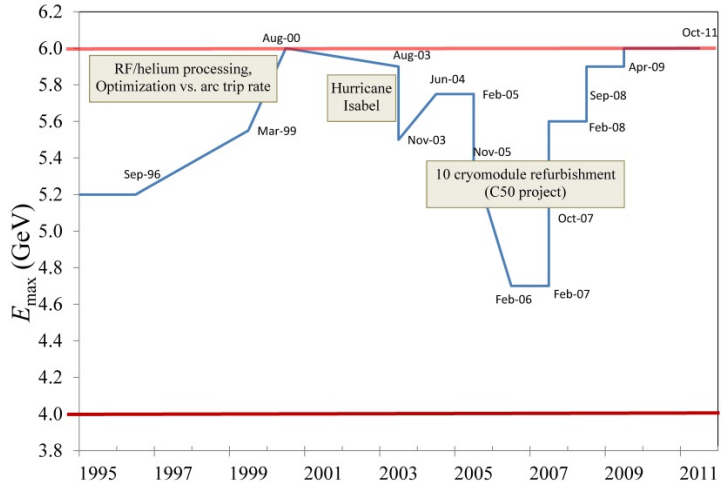

Figure 21. CEBAF maximum beam energy evolution prior to and during the C50 project.

While the C50 cryomodules filled the need for linac voltage to support 6 GeV physics, the RF losses were higher than expected. Although the refreshed cavity pairs performed very nicely in their VTA acceptance tests, $Q_0$ tests in the completed cryomodules showed significant degradation, essentially unchanged to performance prior to refurbishment except for near elimination of the field emission loading contribution. This puzzle is illustrated by Figure 22, which shows the performance of cavity IA-085 as it was installed in CEBAF in 1993, its performance in the VTA in 2007, and its performance reassembled in the C50-4 cryomodule in 2007-8.

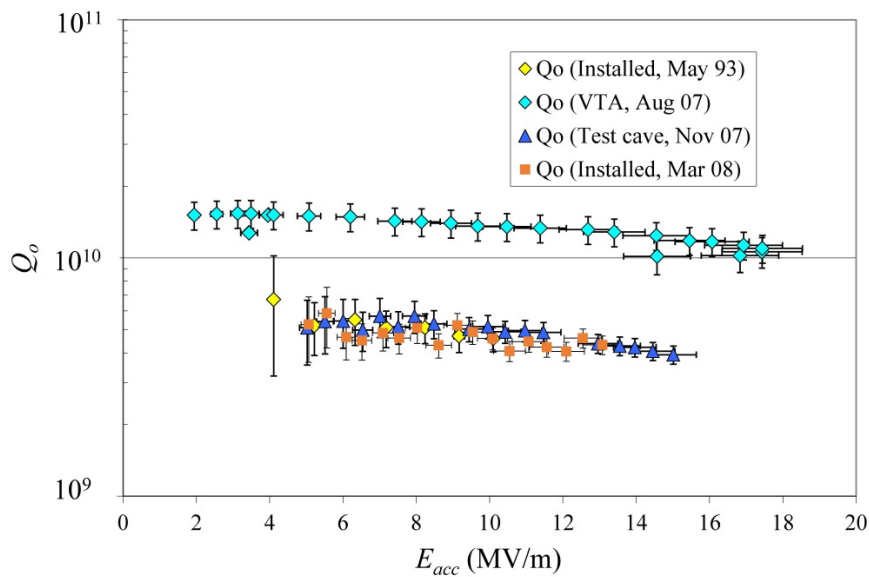

Figure 22. Performance of cavity IA-085 as installed in CEBAF in 1993, in reworked qualification VTA and cryomodule acceptance tests, and as recommissioned in CEBAF in 2008. From [106].

Clearly there existed unexpected loss contributions not related to the basic cavity surface quality. In the middle of the project, efforts were spent to identify unrecognized magnetic field sources. Several associated with the tuner and were found and mitigated, however the problem was not fully eliminated. The project concluded without full resolution of the problem, but did conclude with CEBAF well positioned to support several years of robust physics running at 6 GeV.

## XVI.   DEVELOPMENT PATH FOR THE C100 CRYOMODULE

### 1.   *New cryomodule design conception*

In the late 1990's the JLab Accelerator Division formed an Accelerator Development Department under the direction of Jean Delayen. The mission was to both optimize the capability of the existing CEBAF and to develop designs for an eventual upgrade of CEBAF. The goal was first 10 GeV and then shifted to 12 GeV in response to the emerging physics priorities. Developments in strong-QCD theory indicated that important understanding of quark-gluon behavior, particularly the nature of quark confinement, could be determined by measuring exotic meson spectra with a beam of polarized 9 GeV photons. This beam could be effectively produced with 12 GeV polarized electrons.

By 1999 the conceptual plans had settled on a 12 GeV CEBAF obtained by filling the five empty cryomodule slots in each linac and adding a 10[th] arc and a 6[th] pass through the north linac to provide maximum energy CW electron beam to a new Hall D. This arrangement allows the continued operation of the original three experimental halls in parallel with the new one.

Initially, the building block for the eventual upgrade was envisioned to be new Mk II 70 MV cryomodules used to populate the ten empty zones plus the six weakest existing cryomodules. This architecture was thought to be able to exploit the existing klystrons. Later, after an overall system cost optimization including loss of confidence in pushing the existing klystrons to 8 kW, the goal was shifted to 108 MV cryomodules to populate the ten empty zones and no replacement of existing cryomodules, but requiring new 13 kW klystrons.[95, 107-109]

The new cryomodules required a new cavity design. To maximize the active length within the existing cryomodule footprint, the number of cells per cavity was increased from five to seven.[110] The tight longitudinal spacing required elimination of all bellows between cavities, so that the series of eight 7-cell cavities was attached directly together to form the central string. In order to eliminate all gasket seals directly between beam vacuum and liquid helium, the helium vessels were reduced from enveloping a whole cavity pair as in the original design to enclosing only the seven cells of each cavity. A new tuner mechanism was thus also required.[111-113] The waveguide higher order mode (HOM) couplers and loads were replaced by coaxial couplers adapted from those in use at DESY.

The input waveguide coupler was retained, but modified to eliminate a field asymmetry in the original that imparted an undesirable transverse kick to the beam and to greatly reduce sensitivities to mechanical fabrication tolerances.[114, 115] The nominal input coupler $Q_{ext}$ was increased from $6.6 \times 10^6$ to $2.2 \times 10^7$. It was also recognized that a more agile low level RF (LLRF) control system would be needed to control the narrower bandwidth of the upgrade cavities. So efforts started early to specify a system which could find, tune, and ramp-to-gradient individual cavities in self-excited loop mode, then transition to fully synchronized regulation mode.[116, 117]

A warm space frame using support rods configured in a double-paired cross pattern at each end was chosen to support each cavity in the cavity string. The space frame is supported inside the vacuum vessel at the quarter points, each being between the second and third cavity from the end. The frame is made up of a series of hoop rings installed perpendicular to the beam axis down the length of the cavity string and connected by axial support members. The frame is assembled around the string along with the tuner assemblies starting from one end and progressing to the other. When the space frame is completed, magnetic shielding, thermal shielding, and associated components were added to the structure. Then the integrated assembly could be wheeled into the vacuum vessel and locked into position. A cross-section of the Mk II cryomodule design is shown in Figure 23.

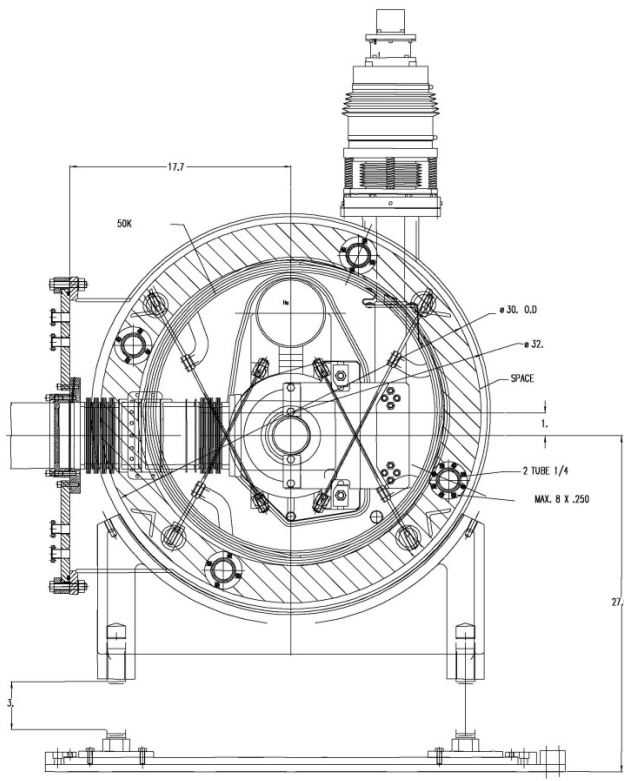

Figure 23. Cross-section of Mk II CEBAF cryomodule.

## 2. *Two C70 prototype cryomodules*

Two cryomodules were constructed based on the Mk II 70 MV specification in 2002-2003.[118, 119] The cell geometry in the 16 cavities in these cryomodules was the same as that in the original CEBAF 5-cell cavities. One of these cryomodules was constructed for the 10 kW upgrade to JLab's Free Electron Laser. It incorporated a few minor modifications to accommodate the significantly higher HOM power generated by the intense short bunches in that accelerator.

The cavities for these cryomodules were fabricated in-house at Jefferson Lab. The final surface treatment was a 20 µm etch followed by 2 hr UPW HPR, final probe assembly, and an additional 2 hr UPW HPR.

Cavity performance in the CMTF was excellent. The average accelerating gradient achieved was 16.8 MV/m, implying a total potential CM voltage of 83 MV. (Figure 24.) Performance realized after installation in CEBAF was considerably lower, only 55 MV. This limitation was due to two factors: the old LLRF system was used and could not effectively provide stable regulation and tune control above 11 MV/m, and the limited superfluid heat flux through the individual cavity riser pipes between helium vessel and the two-phase header was inadequate for operation with RF heat plus heater heat. The later factor was aggravated by the choice made at the time to operate CEBAF at 2.09 K rather than the 2.04 K design condition. 65 MV operation was demonstrated without cryogenic limitation during a subsequent test at 2.04 K.

The diameter of the helium riser pipe in the second Mk II cryomodule was doubled. That module was commissioned in the FEL and demonstrated stable 82 MV operation with 8 kW klystrons without beam.

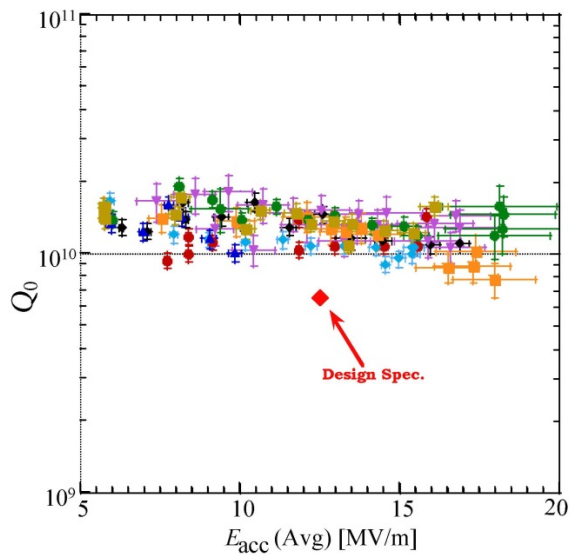

Figure 24. Acceptance test performance of cavities in the first intermediate prototype upgrade cryomodule (SL21) in the CMTF. From [118].

### 3. First C100 prototype cryomodule **Renascence**

In preparation for the CEBAF 12 GeV upgrade, a project to design and build a developmental prototype was launched as a performance improvement effort. This cryomodule was to demonstrate 108 MV acceleration capability with dynamic heat load less than 250 W at 2.07 K, the specifications for the Upgrade. As this module was intended to be a fresh new start incorporating several new improved systems based on the previous prototypes and JLab experience designing and constructing the SNS cryomodules, it was given the name *Renascence*.[120, 121] A cross-section of a cavity and its attachments in this design is shown in Figure 25.

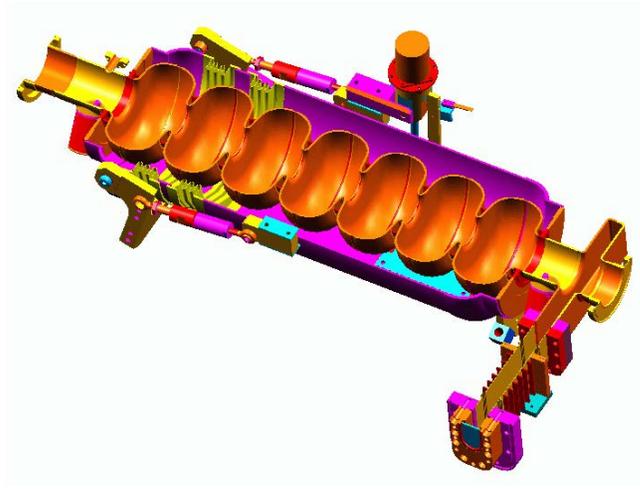

Figure 25. Cross-section *of Renascence* cavity, couplers, helium vessel, and tuner.

This cryomodule included a mixture of two different cell shapes in its eight 7-cell cavities, a new "high gradient" (HG) cell design and a new "low loss" (LL) cell design. The former was an optimization to minimize peak surface electric fields, while the latter was an optimization to attain the greatest voltage/cryo-watt.[122] All of the *Renascence* cavities were fabricated in-house at Jefferson Lab. After fabrication, the cavity surfaces were prepared by BCP etch and hot UPW rinse, then HPR and assembly in Class 10 cleanroom environment. The cavities met the 19.2 MV/m goal with less than 29 W heat in 2.07 K vertical tests.[123] The performance of the LL cavities during qualification tests are shown in Figure 26.

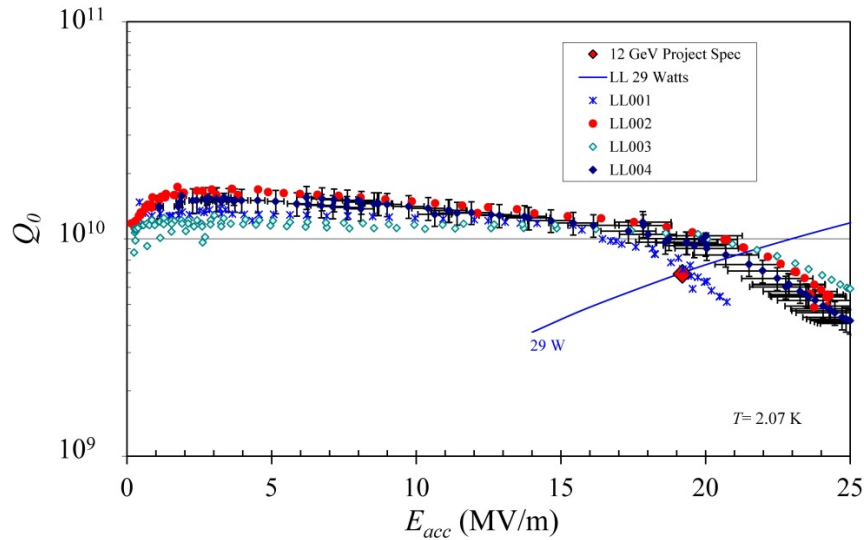

Figure 26. Qualification tests of the four Low Loss (LL) cavities for *Renascence*. From [123].

The HOM couplers were moved closer to the end cells to compensate for the weaker cell-to-cell coupling of the new cavity shapes relative to the original CEBAF shape. Also, two HOM couplers were placed on each end of the cavity to assure adequate damping not only for CEBAF, but also for potential utility in the JLab FEL. As is now typical, the HOM couplers function as broadband couplers with a resonant notch filter rejecting the fundamental accelerating mode. This implies the presence of fundamental mode fields that do not couple out to the transmission line and load. The RF pickup probe in these couplers had to be made superconducting to avoid significant heating on the previous copper probes. [124] This in turn created the need for a custom high thermal conductivity feedthrough to provide a cooling conduction path for the niobium pickup probe This was subsequently developed and patented.[125, 126]

Because of the very limited space constraints between the cavities, a new "Radial-Wedge" clamp was developed and patented for use on all beamline flanges in order to provide high sealing forces and easy assembly.[127] Another unique solution developed was the "serpentine gasket" used to provide a clean UHV cryogenic seal for the rectangular flange of the input RF waveguide.[128]

The *Renascence* cryomodule used a new "nutcracker-style" tuner actuated by a cold stepper motor with harmonic drive and piezoelectric element.[113] This was attractive because of its generous stroke and its avoidance of the need for flexible penetrations through the insulation and vacuum vessel to external mechanical drivers. Issues that arose in use, however, led the cryomodule designers to revert to the scissor-jack style tuner for the Upgrade Project cryomodules.

Assembly of *Renascence* took place in 2005. Initial testing encountered serious problems. Focus shifted to careful diagnosis of the problems and their remediation.[129, 130] Non-standard cryogenic instrumentation feedthroughs developed leaks on initial cooldown below 70 K. Mechanical binding of tuners was a problem. One warm ceramic window (on cavity 5) was found to leak. The most significant problem, however, was end group quenching of all cavities at low gradient during CW testing. With 10% duty cycle pulsed RF operation, the cavities performed fine to 17-21 MV/m. Investigation revealed that the high thermal conductivity RF feedthroughs in the HOM couplers were themselves not provided with adequate thermal anchors, with the result that during sustained operation the temperatures of the superconducting Nb probes would drift up above $T_C$, at which point the cascade of increasing dissipation would lead to cavity quench.

Since HOM damping measurements accomplished during initial testing demonstrated the adequacy of using only two HOM couplers, the decision was taken to remove the two unnecessary coupling probes which were also placed in the most vulnerable thermal location. In principle, this probe removal could have been accomplished without full disassembly of the cavity string. To minimize the risk of particulate contamination, however, the decision was taken to fully disassemble the string and High Pressure Rinse (HPR) each cavity again following probe removals. Upon reassembly, thermal strapping was added to the remaining HOM probe feedthroughs and resulted in excellent thermal stability under high field CW operation.

After reassembly, six of the eight cavities in *Renascence* performed well. One with very low $Q_0$ was found later to have been damaged during its HPR cleaning prior to reassembly, and the tuner of another cavity was found to be stuck and unable to reach 1.497 GHz required for use in CEBAF. The cryomodule was installed in CEBAF zone NL04 in August 2007. With the available 6.5 kW RF, the six operable cavities could be run at 14 MV/m, providing 59 MV. Having already been the vehicle for debugging several systems that proved vital for the CEBAF Upgrade, *Renascence* had yet one more lesson to reveal.

Recirculating accelerators can be vulnerable to a particular type of beam instability in which a slightly off-axis beam stimulates a dipole mode in an accelerating cavity sufficiently to affect the same bunch when it returns on a subsequent pass. This is called multipass beam break-up (BBU). Superconducting RF cavities, with potentially long high-$Q$ memories require damping of such dipole modes in order to avoid such instabilities. Although previously actively pursued during the initial design analysis for CEBAF, which demonstrated that the 5-cell cavity design was BBU-stable up to tens of mA [131], and experimentally verified during the front end test [57], BBU was first encountered in CEBAF with *Renascence* in the fall of 2007.[132] A beam current limitation as low as 40 µA with an oscillation frequency of 2.156 GHz was traced to cavity 5 (HG002) in this cryomodule. The accelerator beam physicists were delighted to have an opportunity to test their BBU model with different beam orbits and supplemental damping added to the cavity. Two dipole modes in the HG cavity in position 5 of *Renascence* were found to be very weakly damped, with $Q_l \sim 1 \times 10^8$. The investigation and characterization process provided a rare experimental test of the multipass BBU beam dynamics theory, building on the experience with the JLab Free Electron Laser reported in 2005.[133] The beam and RF test results were consistent with an abnormally shortened cavity. Subsequent investigation of records in Pansophy revealed that a double weld had been made on a cell equator of this HG cavity during fabrication. Inelastic tuning of the cells required to attain the target fundamental mode frequency resulted in an abnormal field configuration for the two dipole modes which were then poorly damped and created ideal conditions for BBU.[134]

After the BBU experience with cavity HG002, additional quality assurance checks involving loaded-*Q* measurements for all relevant HOMs were instituted as part of the VTA cavity tests for all later CEBAF upgrade cavities. Also, cavity fabrication tolerances were tracked much more closely for JLab-built cavities and subsequent vendor-supplied cavities.

### 4. Final C100 cryomodule design refinements

Well before *Renascence* arrived in CEBAF, the lessons learned motivated several changes to the "C100" cryomodule design for the Upgrade. These included a few significant changes to the cavity assembly.[135] For the C100 cavity, the LL 7-cell structure was maintained, but the stiffening rings were eliminated to ease tuning. The number of HOM couplers was reduced to two and moved further away from the cavity end cell, with ports oriented so as to allow a return to the scissor-jack tuner and addition of a brazed niobium-to-stainless transition for the helium vessel. Substitution of a stainless steel helium vessel for titanium was taken as both a cost saving and reliability improvement.[136] The changes to the HOM couplers reduced the calculated heat loss on the sensitive pick-up probes by a factor of 90. Generous thermal strapping was maintained.

The C100 cavity design optimization and the fabrication and qualification of the first two niobium 7-cell C100 cavities took place in 2006. The cavities were fully fabricated at JLab. The cavities received the current standard BCP/600°C degas bake/tuning/BCP/HPR processing treatments. The cavities were tested individually and assembled together and tested with tuners and waveguide coupling in the Horizontal Test Bed (HTB), in which the thermal boundary conditions replicate those of a full cryomodule. This HTB test was completed in January 2007 and demonstrated all of the performance characteristics required for the 12 GeV Upgrade. This included a thorough HOM survey that found a safety margin of at least an order of magnitude to the corresponding impedance budget.[137]

With the completion of full qualification of the C100 cryomodule design and the associated LLRF controls[138, 139], the acceleration system components of the CEBAF 12 GeV Upgrade project were ready. The scientific motivation for the project was fully established and the required beam transport modifications were fully scoped and ready for DOE approval to proceed with construction.[140]

## XVII.   PRODUCTION AND PERFORMANCE OF C100 SRF CAVITIES

Formal start of the CEBAF 12 GeV Upgrade Project occurred in early 2009. A contract for fabrication of 86 C100 cavities for the project was then awarded in July 2009 to Research Instruments (RI), Germany.  By January 2010, the statement of work was amended to add a post-production chemical cleaning of cavities by bulk (160 µm) Buffered Chemical Polishing (BCP) etch with subsequent final tuning.[141] All 86 cavities were delivered successively between July 2010 and March 2011. Per specification, the cavities were tuned to a warm target frequency (1494 MHz ± 100 kHz). All cavities were fabricated from RRR > 250 fine grain niobium sheets supplied by Tokyo Denkai using standard deep drawing and electron beam welding techniques.

In parallel with the Upgrade project cavity procurement, eight "R100" cavities, identical in design to the C100 cavities, were fabricated in-house at JLab in 2010 in order to provide an additional cryomodule (*R100*) eventually needed in the CEBAF injector during the 12 GeV era. This set of cavities was fabricated to a new tighter standards in order to establish high confidence in HOM configuration.[142] The full cryomodule was completed by April 2011.

Contemporaneous with the preparations for the 12 GeV project, JLab had been engaged in construction of cryomodules for the Spallation Neutron Source (SNS) at Oak Ridge National Laboratory and also in the International Linear Collider (ILC) R&D program. The ILC requirement of very high accelerating gradients for that pulsed SRF application provided support for R&D work on processing techniques that included improved electropolishing (EP) of niobium cavity surfaces. [143, 144] The microscopically smoother surface is associated with higher $Q_0$ (lower RF losses) at higher sustained RF fields. As these techniques and understandings were tested on CEBAF 7-cell cavities, it became clear that the 12 GeV Project would benefit from adopting an electropolish as the final surface processing step.[145, 146] Figure 27 shows the excellent performance of the first twelve 7-cell, in-house built LL cavities that received a light electropolish as the finishing step, including all of the R100 cavities.

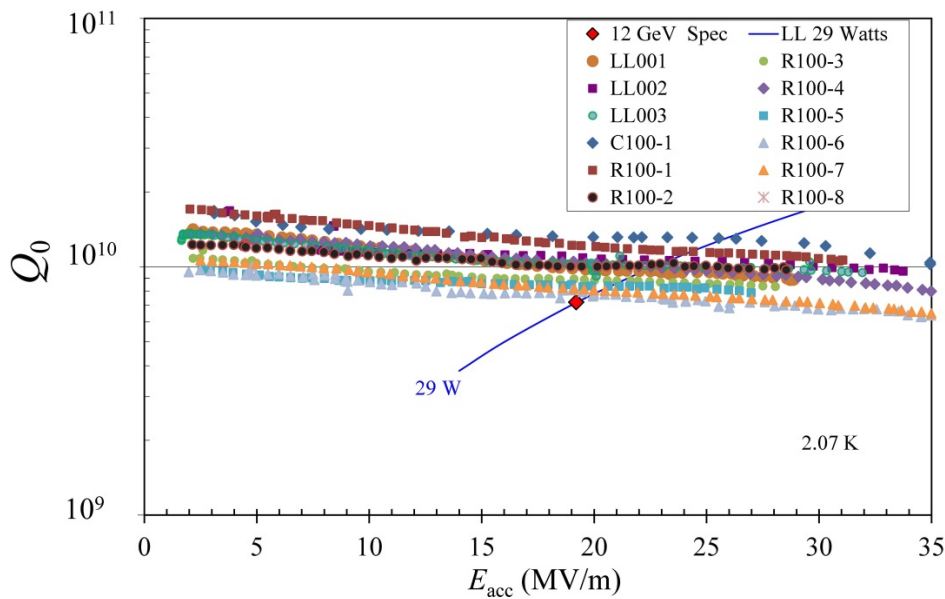

Figure 27. Performance of CEBAF 7-cell cavities after ~30 μm EP. From [146].

Although the 12 GeV Upgrade project baseline called for cavities prepared only by BCP etch, the excellent results from parallel process development enabled the project to adopt a modification calling for a final 30 μm EP followed by 24 hour bake at 120 C just prior to testing was implemented by JLab as a performance risk reduction measure.

Having had plenty of process development time prior to the arrival of the production stream of cavities and the excellent performance of the cavity vendor, the 12 GeV cavity production line ran very smoothly.[147-150] The cavity performance during VTA testing significantly exceeded requirements such that most of the cavities were not actually tested to their limits, but were only tested to an administratively constrained 27 MV/m. The electropolishing process and cavity performance was so stable and reliable that the decision was made for efficiency to only test the cavities after the helium vessels were welded on. One early production cavity that was tested to its limits was C100-6. Its excellent performance is illustrated in Figure 28. Subsequently, after the addition of the helium vessel around a cavity, the maximum cooling capacity at the 2.07 K test temperature was ~70 W.

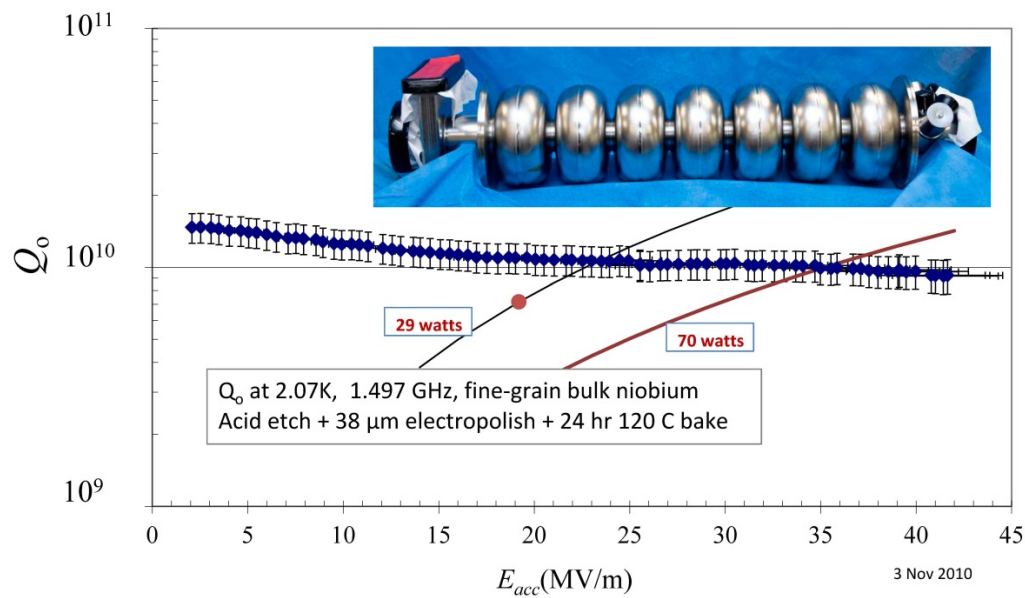

Figure 28. Performance test of cavity C100-6 without helium vessel.

## XVIII.  C100 CRYOMODULE PRODUCTION AND TESTING

The eight C100 cavities that make up a string for the CEBAF Upgrade cryomodules are bolted directly together without intervening bellows. As mentioned earlier, this design choice was driven by the existing tunnel space constraints. Following qualification RF tests, the cavities were accumulated in the cleanroom.  Each cavity was given a final UPW HPR rinse, waveguide window assemblies were attached, and the cavities were sequentially assembled on an assembly rail in the Class 100 area. See Figure 29.

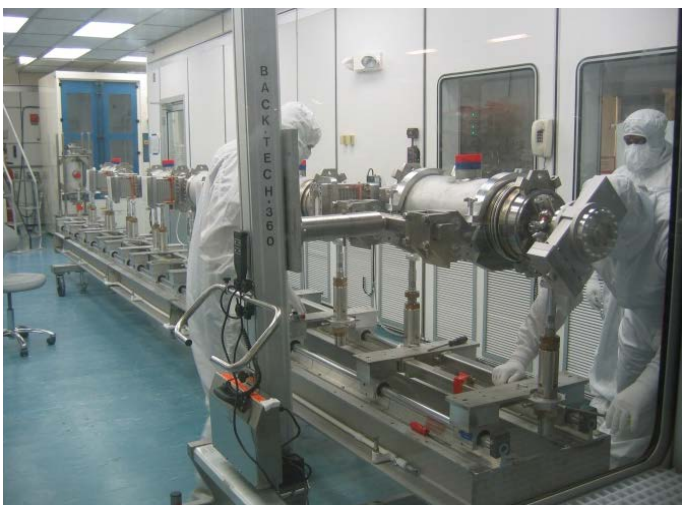

Figure 29. C100 cavity string assembly.

After evacuation of the string, it was rolled out of the cleanroom and handed off to the cryomodule assembly crew who established final alignment, instrumentation, helium plumbing, insulation, structural support and final vacuum.[151]

Acceptance testing of the first cryomodule, C100-1, began in the cryomodule test facility (CMTF) in June 2011. This cryomodule and also C100-2 were installed early in CEBAF and operated with beam during the final six-month run of the 6 GeV era, November 2011 – May 2012. At the end of this period C100-2 successfully demonstrated the full 108 MV performance with the 465 µA design current.[152]

The first cryomodules met all of the performance requirements, but the observed sensitivity to microphonics was more than anticipated. Because of the cavity's very narrow band resonance (47 Hz) with $Q_1$ of $3.2 \times 10^7$, even relatively minor detuning effects quickly place a heavy demand on the RF control power required to regulate the cavity fields. A careful investigation yielded both a solid explanation and an effective solution to the excessive microphonics.[153] The combination of directly-connected cavities and the removal of the stiffening rings on the LL-cell-shaped C100 cavities provided an opportunity for low-frequency ridged-body vibrational modes of the whole string (~10.5 Hz) to drive coherent microphonic effects on all cavities in a cryomodule with an amplitude of ~12 Hz peak detuning in the CEBAF tunnel.

It was found that increasing the stiffness of the cavity tuner was sufficient to solve the problem. The change was implemented on cryomodules C100-4 through C100-10 and produced an average reduction in microphonics of 42%.

By September 2013, all of the C100 cryomodules were installed in CEBAF. Their in-place SRF checkout tests found integrated voltage performance ranging from 105 MV to 124 MV based in individual cavity gradient maximums.[154, 155]

## XIX.   C100 CRYOMODULE PERFORMANCE IN 12 GEV CEBAF

### 1.   RF controls

The RF system used for the CEBAF Upgrade cryomodules is of a completely new design.[156, 157] Each cavity is powered and controlled by a single klystron and LLRF system. The klystrons produce 12 kW of linear power and up to 13 kW saturated. Each cavity field and resonance control PI algorithm is contained in two FPGAs. One FPGA is in the field control chassis, controlling a single cavity. The resonance control chassis contains the other and controls up to eight cavities. The RF controls are unique, incorporating a digital self-excited loop (SEL) to dynamically manage the RF frequency during turn-on and trip recovery.[155]

Lorentz force effects shift the C100 cavity resonance frequency by ~17 bandwidths (800 Hz) during turn on to ~20 MV/m. Other accelerators have used a piezoelectric tuner to compensate for this by using a feed-forward algorithm. JLab went in a different direction and uses the self-excited loop (SEL) to track the cavity frequency as it changes during the fill and quickly restores the cavity to its operational gradient. A firmware application then tunes the cavity and switches to Generator Driven Resonator (GDR) mode, locking the cavity to the reference required for beam operations. Figure 30 shows a plot of the forward power/phase and I/Q signals as it is switched from SEL to GDR. The routine has become automated with a "one-button" turn-on routine, making the high *Q*, high gradient cavities much easier to setup than the older, low-gradient cavities.

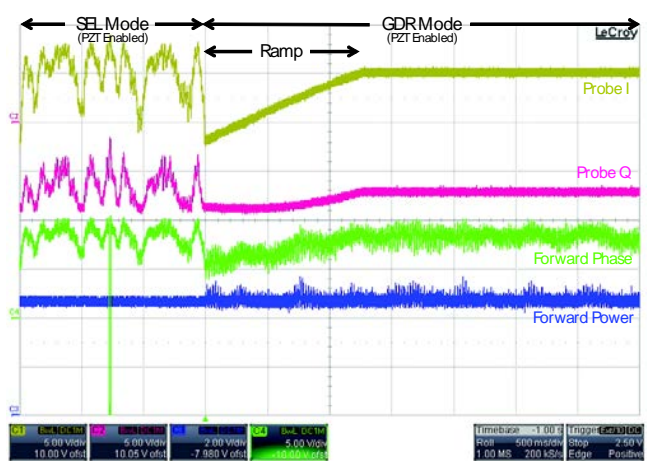

Figure 30. Annotated oscilloscope image capturing the transition from self-excited-loop (SEL) to generator-driven-resonator (GDR) control with piezotuner enabled. From [155].

The direct mechanical coupling of the C100 cavities causes the Lorentz force tuning of a cavity to affect its neighbors. The trip off of one cavity imparts an impulse that detunes the neighbor which risks tripping it as well. To manage this, in the event of a fault, neighboring cavities are switched to SEL mode to easily ride out the transient until the trip is recovered. Then all are switched back to GDR mode for beam operations.

The first three C100 cryomodules were outfitted with piezoelectric fine tuners on each cavity in addition to their mechanical tuner. These worked very well for eliminating static drift detuning while running in GDR mode. (See Figure 31.) They are not useful for compensating microphonics. Because a mechanical tuner control solution was found that works better than expected, piezo tuners were not installed on C100-4 through C100-10.

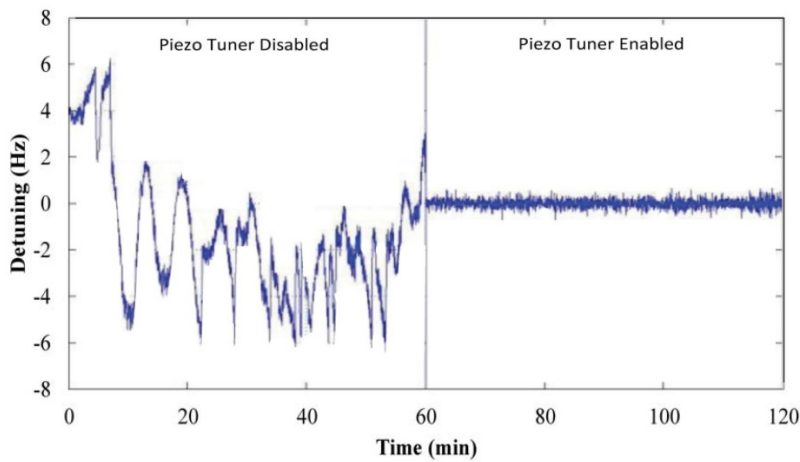

Figure 31. Piezotuner control off and on. From [155].

## 2. Heat management

The diameter of the helium riser pipe extending from the helium vessel of each C100 cavity is designed to conduct up to 65-70 W at 2.07 K. Above this heat flux, pressure and temperature instability will occur. At design specification, each C100 7-cell cavity may dissipate up to 29 W. If, for some reason, the $Q_0$ is low or the cavity is operated at higher than nominal gradient, additional heat may be generated. In addition, a resistive heater is operated within each helium vessel to provide a compensating cryogenic load when the RF is off. (The cryogenic plant requires a stable load.) In the original CEBAF cryomodules, recall that the helium vessels were large, encompassing a whole cavity pair. One heater control was used to drive all of the heat within a cryomodule, distributing it uniformly across four large cryounits. In the C100 cryomodules, the same controller was at first used to distribute the heat among only four of the eight smaller helium vessels. This occasionally caused the total heat within one cavity's helium vessel to exceed the maximum stable amount leading to instabilities and trips. The heater heat was subsequently distributed evenly among all eight cavities. An optimal, more sophisticated heat management system remains yet for future implementation.

## 3. HOM damping checks

After the installation and commissioning of the first two C100 cryomodules in CEBAF, opportunity was taken in November 2011 to accomplish an experimental evaluation of the C100 dipole HOM damping and BBU thresholds.[158] This check was motivated by the earlier encounter with BBU with one of the *Renascence* cavities.[134] The survey confirmed that all of the dipole modes of potential concern were adequately damped. This complemented a very thorough QA tracking of these modes from VTA tests, to CMTF tests, and post-installation tests.[159] Only one mode on one cavity (TM111 $\pi/7$ vertical polarization of C100-9 in cavity 6) exceeded the baseline impedance budget of $2.4 \times 10^{10}$ $\Omega$/m. It could be brought within budget via the use of an external stub tuner. These data provide confidence that CEBAF will be able to operate free of BBU in the 12 GeV era.

## 4. Ready for 12 GeV

The high-level beam parameters delivered by 6 GeV CEBAF and expected from 12 GeV CEBAF are listed in Table 4.

Table 4: Delivered Beam Parameters for 6 GeV CEBAF and Expected Beam Parameters for 12 GeV CEBAF. (Reproduced from [160].)

| Parameter | 6 GeV | 12 GeV |
|---|---|---|
| Max. Energy ABC | 6 GeV | 11 GeV |
| Max. Energy D | -- | 12 GeV |
| Duty Factor | CW | CW |
| Max. Beam Power | 1 MW | 1 MW |
| Bunch Charge (Min-Max) | 0.004 fC – 1.3 pC | 0.004 fC – 1.3 pC |
| Hall Repetition Rate (Min- | 31.2 – 499 MHz | 31.2 – 499 MHz |
| Nominal Hall Repetition Rate | 499 MHz | 249.5/499 MHz |
| Number of Exp. Halls | 3 | 4 |
| Max. Number of Passes | 5 | 5.5 |
| Emittance (geometric) at full | 0.1 nm-rad(X)/0.1 nm- | 3 nm-rad(X)/1 nm- |
| Energy Spread at full energy | 0.002% | 0.02% |
| Polarization | 35%(initial), 85%(final) | >85% |

All C100 cryomodules were operated for the first time from January 2014 through May 2014. The initial commissioning goals of 2.2 GeV/pass which corresponds to 12 GeV in the machine and injector energy of 123 MeV were achieved.[161] During this time the new cryomodules also supported preliminary beam delivery to the experimental halls. Cryomodule voltages ranged from 50 MV to over 100 MV, depending on the requirements for the testing.[162] The program goals for the accelerator portion of the 12 GeV Upgrade Project were achieved five months ahead of schedule.[160] Beam was delivered simultaneously to Halls A (4-pass), B (1-pass), and D (5.5-pass) during the late 2014 run period.

In October 2015 a program was run for two weeks to accumulate fault data sufficient to create new models for the linac energy management system (lem).[71, 76] Cavities were ramped up until faults occurred at intervals under 10 minutes, held until five true arc faults accumulated, then ramped down 0.5 MV/m. Accumulation repeated in this fashion; last step down typically had fault intervals of one day, twice the operating rate desired.

After fault models were developed and used, cavity momentum gains were recalibrated using the arcs as spectrometers. A novel phase shift method replaced the gradient balance method previously used. The beam was offset ~5 mm in the arc, the BPM limit for good accuracy. Each cavity was then offset six negative increments in phase such that the beam moved up to 10 mm inboard from original the location. Another six positive phase increments were then made. Phase offsets ranged to 165 degrees so almost twice the perceived momentum change of the cavity could be applied, a much larger change than with a momentum balance approach. At each phase offset the momentum of the beam in the arc was calculated. Subsequently the twelve points were fitted to provide actual momentum and phase offset of the cavity. The new calibration factors were downloaded in early December 2015. $Q_0$ and fault models were adjusted in concert. [163]

The highlight of FY16 for CEBAF Operations was the delivery of 70 µA at 11 GeV to Hall A dump about 22:00 on December 15, 2015. Current was administratively limited, not hardware limited. Beam was subsequently RF separated to allow CW delivery to Hall A at fifth pass and Hall D after a sixth trip through the north linac, yielding a bit over 12 GeV after synchrotron radiation losses. During spring 2016 CW beam was delivered at 12 GeV to Hall D and three different energies to Hall A, for commissioning and two physics experiments respectively.

### 5. C100 cryomodule limitations

Several of the installed C100 cavities were found to show signs of field emission loading and corresponding x-ray and neutron production when operated at nominal operating gradients. Because the integrated voltages are higher and the clear aperture within the C100 cryomodule is larger than the original design, field emitted electrons from one cavity may rather easily propagate and gain high energy before striking something.[164] The consequences may be local heat deposition, x-rays, and component radioactivation. In a few cases, it was found that by some causal route high levels of field emission in particular cavities induced elevation of the nearby beamline ion pump pressures, leading to delayed trips with about 1-hour periodicity. Once identified, these cavities were operationally derated. Efforts were initiated to identify the source of the presumed contamination that results in field emission in the cavities and potential remediation methods.

In addition, it was found that although the quiescent microphonic response met requirements in the CEBAF tunnel environment, mechanical couplings to the outside world transmitted occasional disturbances into the cryomodules inducing excessive tuning transients by, for example local truck traffic. Efforts to decouple the vibration transmission into the cryomodules have been initiated.

## XX. CEBAF 12 GEV OPTIMIZATION AND ON-GOING IMPROVEMENTS

### 1. Facility improvements

In the process of serving the evolving needs of the CEBAF nuclear physics science mission, JLab has continued to develop its technical facilities and staff in ways that also benefit other DOE programs. The 24 cryomodules for the SRF linac at SNS were provided by JLab.[165] Prototyping work and process development work for other DOE labs as well as numerous international partners has also been accomplished at JLab.

In the middle of the CEBAF 12 GeV Upgrade project, Jefferson Lab won a competitive award from the DOE Science Laboratory Infrastructure (SLI) program, to build the Technology and Engineering Development Facility Project (TEDF). Among other things, this included an entirely new 3,100 $m^2$ purpose-built SRF technical work facility that was occupied in summer of 2012. All SRF work processes with the exception of cryogenic testing have been relocated into the new building. All cavity fabrication, processing, thermal treatment, chemistry, cleaning, and assembly work is collected conveniently into a new LEED-certified building. An innovatively designed 800 $m^2$ cleanroom/chemroom suite provides long-term flexibility for support of multiple R&D and construction projects as well as continued process evolution.[166]

This 2nd-generation SRF facility was not completed until after assembly of all of the cavity strings for the C100 cryomodules was complete, so the 12 GeV Project was not able to benefit from the infrastructure improvements. The new facility is now in use for continued paced rework of original CEBAF cryomodules, multiple on-going R&D programs and collaborative work, and the testing and assembly of cryomodules for the LCLS-II project.[167, 168] In 2015-2016 a new cryogenic control system was implemented for the SRF 8-dewar vertical test area,[169] replacing the one commissioned in 1991 that served for over 4000 tests. These new facilities offer the prospect of several more decades of service to CEBAF and wider SRF needs.

## 2. Further CEBAF improvements in progress

Current SRF needs for CEBAF that are being addressed include fabrication and assembly of a new low-energy cryomodule for use in the CEBAF injector. This module must accept the initial ~200 keV beam and yield a 5 MeV beam to match into the balance of the injector system. The original "injector quarter cryomodule" design simply used a standard pair of 5-cell cavities. These structures are poorly matched to the low-energy (sub-relativistic) electron beam, and they also suffer from field asymmetries which degrade the beam emittance. These cavities are being replaced by a new 2-cell $\beta$=0.6 cavity and a 7-cell $\beta$=0.97 cavity.[170] This replacement unit will be ready for testing in late 2016.

While CEBAF has successfully delivered 12 GeV beam and the physics program has restarted, work is ongoing to maximize system reliability and robustness. For project cost management, many subsystems have been specified so as to operate adequately, but with very little margin. These low margins eat into reliability and thus physics productivity. Explorations are underway regarding where effort would best be applied to maximize the productivity of 12 GeV era physics.

One proposal that may be cost effective calls for serious re-fabrication of 5-cell cavities from original cryomodules removed from CEBAF for rework. The concept is to exchange the five original Nb cells for a new 5-cell assembly made from improved material and more optimal shape, while retaining the cavity endgroups. If this is accomplished in parallel with resolution of the other residual magnetic field issues, the result may be a recycled cryomodule solution capable of 75 MV with cryogenic load comparable to present 30 MV cryomodules, and free of periodic arc trips. Development work in this direction began in 2016.

As mentioned previously, the "C50" project was successful in gaining the increased voltage required to secure and run the 6 GeV research program, but the cryomodules consistently failed to meet the anticipated $Q_0$ (heat load) target. Ongoing support for CEBAF seeks to fully diagnose this issue and mitigate it in future maintenance cryomodule rebuilds, whether of the C50 or potential C75 style. Recent work has identified previously unrecognized sources of $Q$-limiting magnetic fields internal to the cryomodules. High permeability material problems with springs in the tuners and other stainless steel components were found and are being addressed.[171] In the CEBAF cryomodule design, all of these components were inside of the cold magnetic shielding. Fresh attention is also being applied to analysis of potential anomalous RF dissipation in the cold RF windows and their adjoining Cu-plated waveguides.

One phenomenon receiving fresh attention is the observation of gradual degradation of peak usable gradient in some operational cavities, all associated with incremental increase in effects associated with field emission.[77, 160] Since field emission in SRF cavities is almost exclusively attributed to residual surface particulate contamination, fresh scrutiny is being applied to identify and mitigate particulate transport mechanisms on the beamline and also to phase in the higher particulate control standards that have developed in the community over the past 25 years.[172, 173]

With sharpened control over the few degradation effects and sustained incremental improvements via the on-going maintenance program, JLab expects to establish the SRF-based acceleration capacity of CEBAF sufficient to solidly support a robust 12 GeV physics program for the next two or three decades. Planning and design has also begun on SRF structures that would be needed for a proposed next-generation nuclear physics research facility to be added to CEBAF. This electron-ion collider would receive 10 GeV electrons from CEBAF to collide with ions accelerated by a new complex to be located immediately to the north of the present JLab campus.

# XXI.   CEBAF PHYSICS HIGHLIGHTS

## *1.   Summary of physics enabled by the CEBAF CW SRF accelerator*

Jefferson Lab supports one of the largest scientific user communities in the world, with more than 1,500 researchers whose work has resulted in: scientific data from 178 full and 10 partial experiments; 380 Physics Letters and Physical Review Letters publications; and 1,292 publications in other refereed journals to-date.  Collectively, there have been more than 113,000 citations for work done at JLab. Research at Jefferson Lab has contributed to thesis research material for nearly one-third of all U.S. Ph.D.s awarded annually in nuclear physics, with 531 produced to-date.

CEBAF's completed 6 GeV program has given the U.S. leadership in addressing the structure and interactions of nucleons and nuclei in terms of the quarks and gluons of Quantum Chromo Dynamics (QCD), the field theory of the strong force.

## *2.   Recent physics highlights*

Understanding the internal charge and magnetic structure of the proton is a critically important goal of nuclear physics. Measurements performed over the last decade at Jefferson Lab revealed puzzling behavior that required a new experiment to precisely compare scattering of matter and antimatter (electrons and positrons). Results of this new experiment using the CEBAF Large Acceptance Spectrometer (CLAS) indicate that two-photon exchange processed in addition to single-photon exchanges can resolve the previous experimental puzzle.

The description of complex nuclei at a fundamental level must include correlations among the nucleons beyond the simple mean field model. Indeed, these correlations are essential to understand the details of nuclear structure evident in the spectra of all nuclei. In a recent discovery, these correlations were shown to be associated with the distributions of quarks in nuclei. This provides the first indication of a deep connection between the role of nucleon-nucleon interactions and the quark structure of many nucleon systems.

## *3.   Prospects with 12 GeV CEBAF*

Now firmly established, the 12 GeV Upgrade of CEBAF will enable a new experimental program unique in the world with substantial discovery potential for major advances in our understanding of the substructure of the nucleon, the fundamental theory of the strong force, QCD, aspects of nuclear structure relevant to neutron star physics, and assessments of the completeness of the standard model of particle physics. The new capabilities will enable unique 3D mapping of the valence quarks and extend the earlier studies to describe comprehensively the valence quark momentum and spin distributions in nucleons and nuclei. New opportunities to discover heretofore unobserved hadron states predicted by quantum chromodynamics will become available. Higher precision measurements of the weak couplings of elementary particles will be accessible through measurements of parity-violating asymmetries.

All of these rich physics opportunities are made accessible by the CW superconducting RF systems operating in CEBAF.

# XXII.   ACKNOWLEDGMENTS

The work of developing, designing, constructing, commissioning, and operating the SRF systems for CEBAF as it has evolved over the past 30 years has involved the creative and skilled efforts of several hundred people. Standing on these accomplishments, the future for CW electron linacs is yet brighter than the past. The author is grateful for careful reading of drafts of this paper and contributions by Jay Benesch, Joe Preble, and Geoff Krafft. Theresa Foremaster contributed much of the physics highlights section. This paper is based upon work supported by the U.S. Department of Energy, Office of Science, Office of Nuclear Physics under contract DE-AC05-06OR23177.